\documentclass{article}

\usepackage{arxiv}

\usepackage[utf8]{inputenc} 
\usepackage[T1]{fontenc}    
\usepackage{hyperref}       
\usepackage{url}            
\usepackage{booktabs}       
\usepackage{amsfonts}       
\usepackage{nicefrac}       
\usepackage{microtype}      
\usepackage{lipsum}		
\usepackage{graphicx}
\usepackage{natbib}
\usepackage{doi}
\usepackage{caption}
\usepackage{subcaption}
\usepackage{algorithm}
\usepackage{algorithmic}
\usepackage{amsmath}

\title{Compositing with 2D Vector Fields \\by using Shape Maps that can represent \\Inconsistent, Impossible, and Incoherent Shapes}

\author{ \href{https://orcid.org/0000-0003-3618-4166}{\includegraphics[scale=0.06]{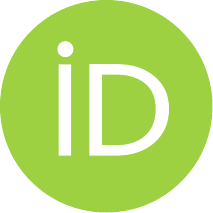}\hspace{1mm}Ergun Akleman}\thanks{Joint with Computer Science and Engineering Department.} \\
	Visual Computing \& Computational Media,\\ Texas A\&M University, College Station, TX, 77831\\
	\texttt{ergun@tamu.edu} \\
       \And
      Youyou Wang\\
Department of Computer Science and Engineering\\ 
Texas A\&M University, College Station, TX, 77831\\
	\texttt{kingyy2010@gmail.com} \\
      \And
	\"Ozg\"ur G\"onen\\
Department of Architecture\\ 
Texas A\&M University, College Station, TX, 77831\\
	\texttt{ozgur.gonen@gmail.com} \\
}

\renewcommand{\shorttitle}{Compositing with 2D  Vector Fields}

\hypersetup{
pdftitle={A template for the arxiv style},
pdfsubject={q-bio.NC, q-bio.QM},
pdfauthor={David S.~Hippocampus, Elias D.~Striatum},
pdfkeywords={First keyword, Second keyword, More},
}

\begin{document}
\maketitle

\begin{abstract}

In this paper, we present a new compositing approach to obtain stylized reflections and refractions with a simple control. Our approach does not require any mask or separate 3D rendering. Moreover, only one additional image is sufficient to obtain a composited image with convincing qualitative reflection and refraction effects. We have also developed linearized methods that are easy to compute. Although these methods do not directly correspond to the underlying physical phenomena of reflection and refraction, they can provide results that are visually similar to realistic 3D rendering. The main advantage of this approach is the ability to treat images as ``mock-3D'' shapes that can be inserted into any digital paint system without any significant structural change. The core of our approach is the shape map, which encodes 2D shape and thickness information for all visible points of an image of a shape. This information does not have to be complete or consistent to obtain interesting composites.  In particular, the shape maps allow us to represent impossible and incoherent shapes with 2D non-conservative vector fields. 

\end{abstract}  

\section{Introduction and Motivation}

\begin{figure}[htbp!]
  \centering
\begin{tabular}{cccccccc}
\includegraphics[width=0.240\textwidth]{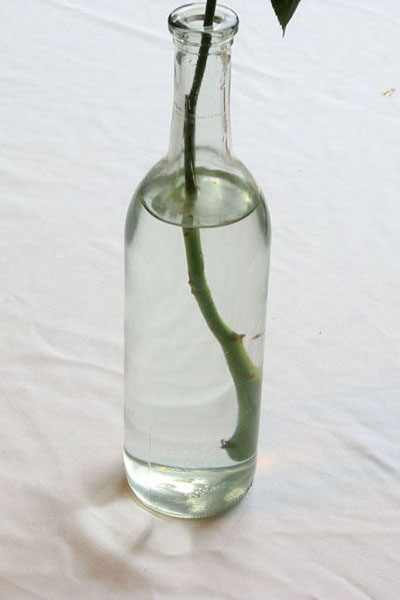}&
\includegraphics[width=0.240\textwidth]{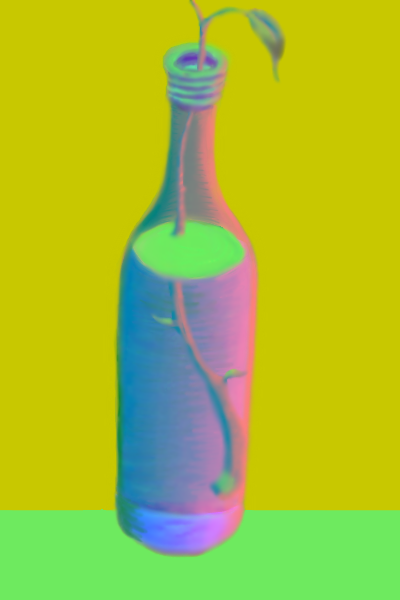}&
\includegraphics[width=0.240\textwidth]{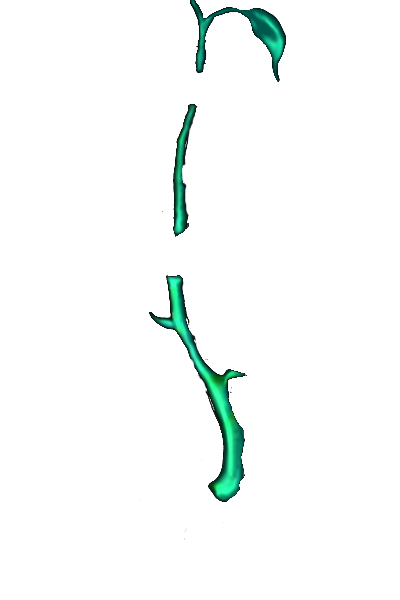}&
\includegraphics[width=0.240\textwidth]{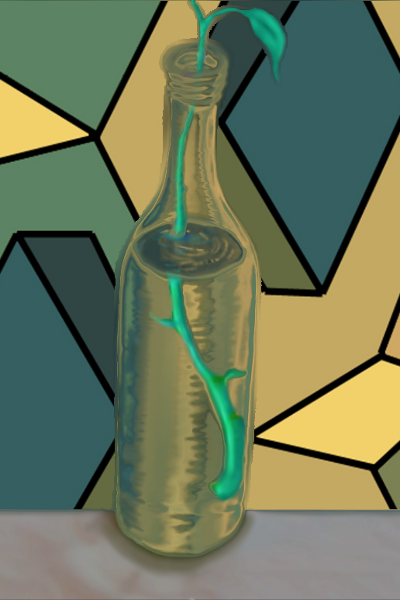}\\
(a) Motivating Photo& (b) Shape Map & (c) Foreground Image $FI$  & (d) A compositing\\
\end{tabular}
\begin{tabular}{cccccccc}
\includegraphics[width=0.240\textwidth]{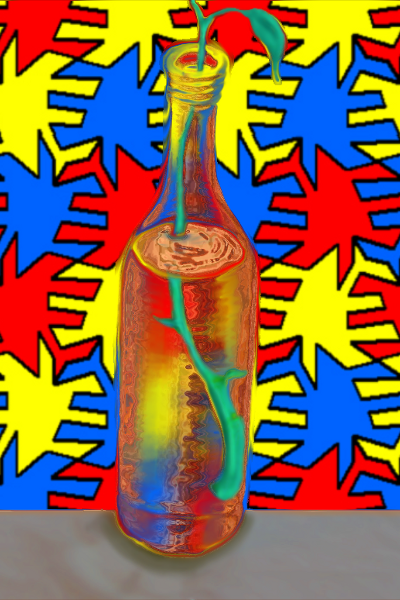}&
\includegraphics[width=0.240\textwidth]{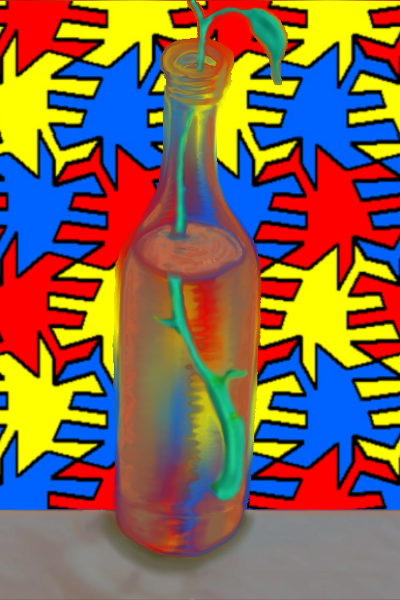}&
\includegraphics[width=0.240\textwidth]{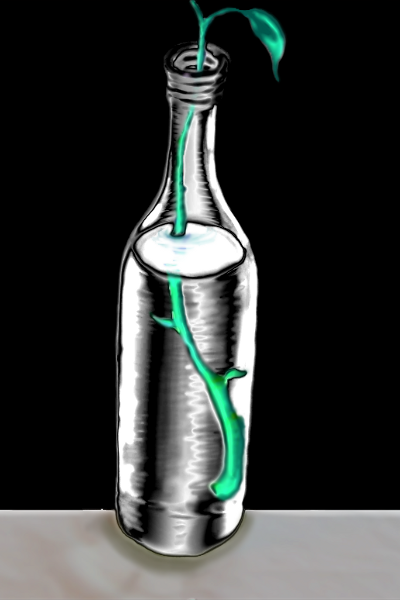}&
\includegraphics[width=0.240\textwidth]{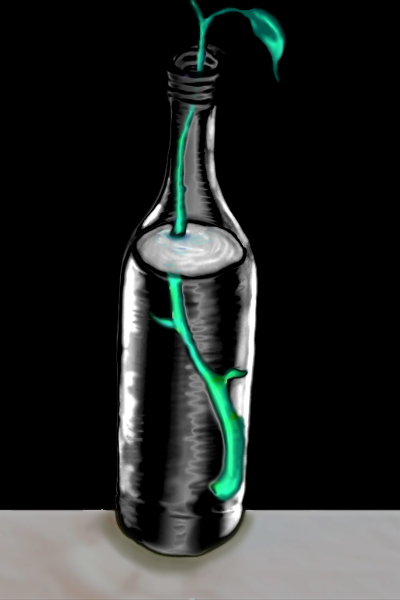}\\
 (e) Refraction and Reflection & (f) Translucent and Glossy  & \multicolumn{2}{c}{(g) Fresnel Effect controlled by pseudo index of refraction} \\
\end{tabular}
\caption{\it An example of compositing with 2D  vector fields by using shape maps.  The particular shape map in (b) is inspired by a photograph of a real bottle in (a). This example demonstrates that a single image included in a 2D digital painting program can be sufficient to obtain a wide variety of refraction and reflection effects. (d) Reflection and refraction combined with the Fresnel term. (f) Glossy reflection and translucent refraction combined with a Fresnel. (f) directly shows Fresnel effects using a black background and a white environment map. }
\label{fig_Bottle}
\end{figure}

Composition is probably one of the most used operations in computer graphics practice \cite{brinkmann2008art}. Let $CI, FI$, and $BI$ denote composited, foreground and background image layers, respectively, and let $\alpha_G$ and $\alpha_{FI}$ denote the global and local opacity of the foreground layer, and the total opacity of the foreground layer is given as $\alpha= \alpha_G \alpha_{FI}$. Then the classical compositing operation is given as:
$$CI = \alpha  H(FI,BI) + (1-\alpha) BI.$$
where $H(FI, BI)$ is a binary operation that defines the type of compositing. The most commonly used operation is over (also called normal) given by $H(FI, BI)=FI$ \cite{Drebin1988}. The binary operations $H(FI, BI)=FI \times BI$ and $H(FI, BI)=FI + BI$ give the other two commonly used operations, namely multiplication and addition. The main advantage of the compositing process is that it provides complete control to 2D artists for image synthesis. Using compositing, an artist can combine a large number of images (i.e. layers) by changing their global alpha's, $H()$ functions, and even images themselves.

\begin{figure*}[ht]
\begin{center}
\begin{tabular}{cccccccccccc}
\includegraphics[width=0.180\textwidth]{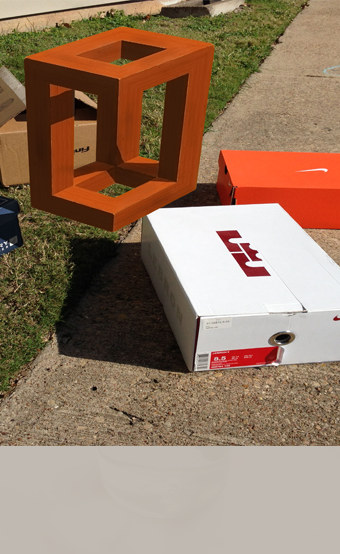}&
\includegraphics[width=0.180\textwidth]{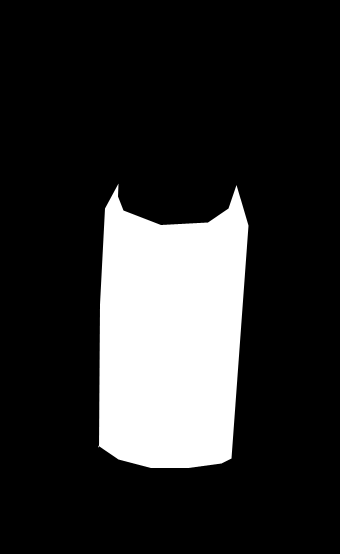}&
\includegraphics[width=0.180\textwidth]{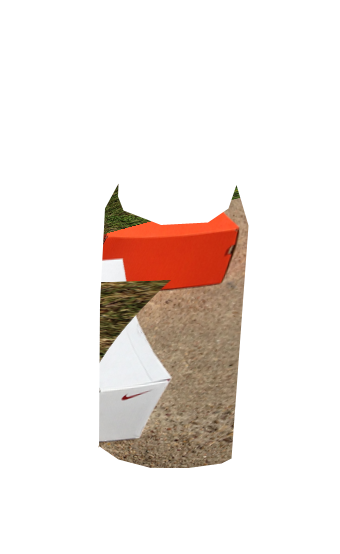}&
\includegraphics[width=0.180\textwidth]{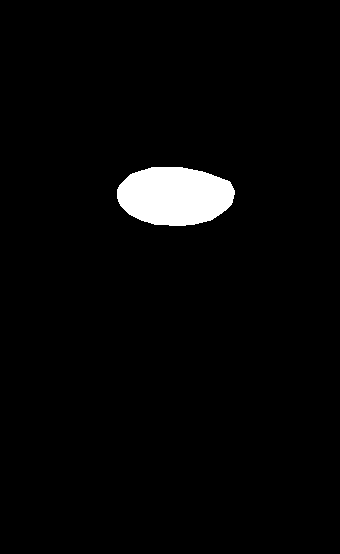}&
\includegraphics[width=0.180\textwidth]{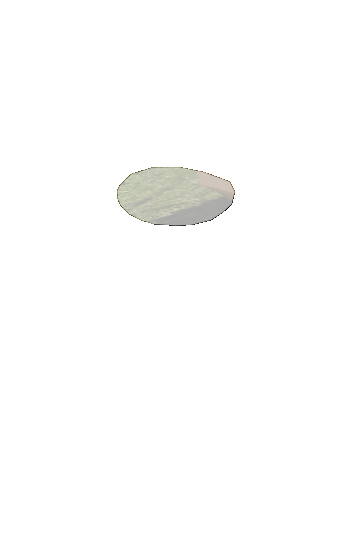}\\
(a) Background& \multicolumn{2}{c}{(b) Water Filled Region Mask \& Refraction} & \multicolumn{2}{c}{(c) Water surface Mask \& Refraction}\\ 
\includegraphics[width=0.180\textwidth]{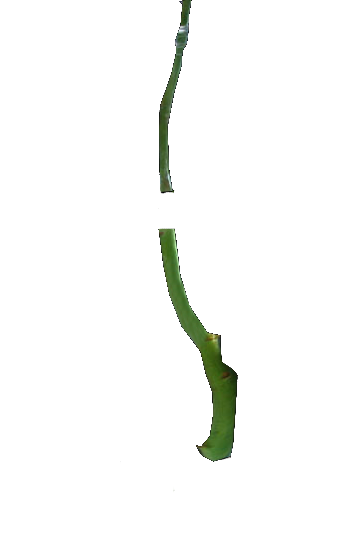}&
\includegraphics[width=0.180\textwidth]{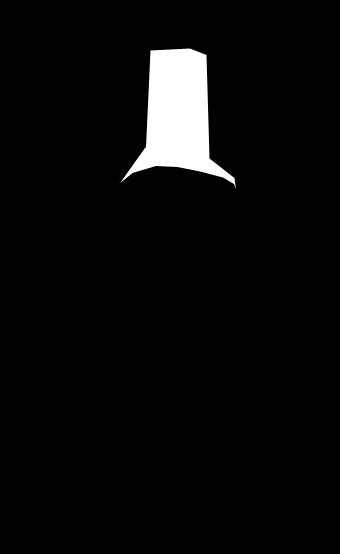}&
\includegraphics[width=0.180\textwidth]{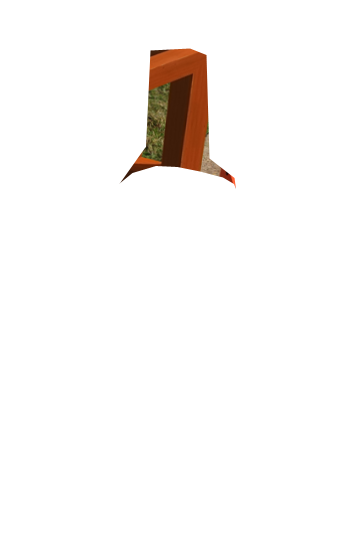}&
\includegraphics[width=0.180\textwidth]{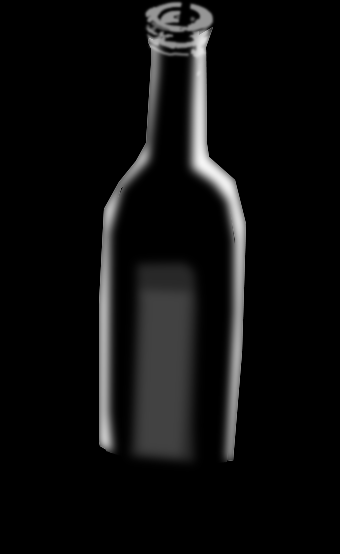}&
\includegraphics[width=0.180\textwidth]{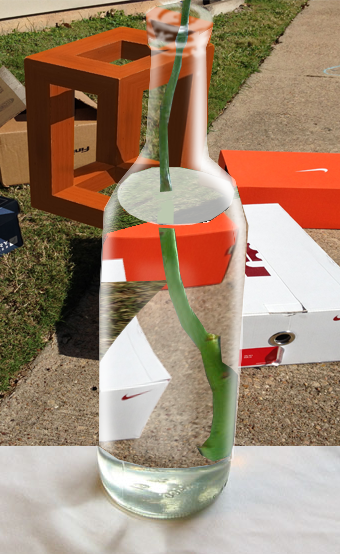}\\
(d) foregrounded & \multicolumn{2}{c}{(e) Neck of bottle Mask \& Refraction }&(f) Fresnel&(g)Final Composite\\
\end{tabular}
\end{center}
\caption{\it Some of the layers that are needed to obtain a final compositing image. }
\label{fig_Bottle2}
\end{figure*}

One of the main issues with this formula is that reflections and refractions are not embedded in the equation. The current 2D digital painting systems do not provide a simple mechanism for obtaining reflection and refraction effects. In current practice, the deformations caused by reflection and refraction must be (1) calculated in a separate 3D rendering software, (2) imported into the 2D digital painting system, and (3) combined using masks (see Figure~\ref{fig_Bottle2}). There is a need for a new compositing approach that can support reflection and refraction effects in the compositing stage.

In this paper, we introduce a compositing approach that provides intuitive control for reflection and refraction by eliminating the need for 3D rendering.  Our new compositing approach uses a shape representation, called ``shape maps'', to compute reflection and refraction effects \cite{wang2014global,wang2014qualitative}. Shape maps are flexible and easy-to-create  mock-3D shape representations. These maps provide efficiency and the ability for 2D artists to include refraction and refraction in their artworks and allow intuitive artistic control over visual results (see Figure~\ref{fig_Bottle}). With shape maps, artists can create artificial, but still believable, versions of the original images, as well as original artwork, that can be dynamically manipulated.

\subsection{Basis and Rationale}
\label{sec_BasisAndRationale}

To demonstrate the need for a new compositing approach, we asked an artist to create a composited image with refraction and reflection using only tools available in a 2D digital painting system. Inspired by the bottle photograph in Figure~\ref{fig_Bottle}, the artist created an image by compositing the bottle with an arbitrary background. Figure~\ref{fig_Bottle2} shows some of the layers the artist created to obtain the final composite image shown in the last column. In this case, the artist used more than 10 layers to handle the different parts of the bottle that refract the background differently. For example, in the water-filled region of the bottle shown in (b), the background will be deformed much more than in the thin glass region in (e) because of the different thicknesses. Moreover, the water surface in (c) can demonstrate a strong reflection due to Fresnel. To handle these differences, the artist created a variety of deformations in the background. Then each deformed image is masked using masks shown in (c), (d), and (e). The artist used additional masks for thick glass regions. In the end, a hand-painted Fresnel term is added to improve realism.

This particular experiment demonstrates the importance of the qualitative nature of refractions and reflections. Although none of the refractions in this image are correct, the composited image is still somewhat acceptable. This is mainly because refractions are qualitatively acceptable. That is, thicker regions introduce more deformation than thinner regions. Another important observation is that the Fresnel term significantly helps the quality of the final result and visually describes the shape of the object.

\subsection{Context and Motivation}
\label{sec_ContextandMotivation} 

Computing refraction and reflection using 3D rendering software can, of course, improve the realism. Fresnel layer can also be computed more accurately using an environment map or light. Although 3D rendering helps produce more accurate composited images, we still need many layers since it is still useful to have masks to avoid registration problems. In other words, masks are needed to make the boundaries of transformed 3D realistic shapes exactly match the boundaries of photographed objects \footnote{Even if the object has a relatively simple 3D shape, we still do not know actual transformations that turn them into images. Furthermore, physical objects can have a wide variety of ``shape imperfections'', which cannot be easily captured with 3D models. In other words, to replace any photographic image of a physical object with a virtual object, registration always presents a challenge}. This can be done by rendering a slightly larger 3D shape and masking the rendered image to have perfect registration. In this process, it is also useful to decompose the original object into a set of simple shapes. For example, the water-filled region in the bottle can be represented by a cylinder and the water surface can be represented by a plane. In other words, in practice, we may again end up with many layers. Moreover, 3D rendering is not an option for 2D artists who do not want to deal with 3D rendering software.

The presented approach is closely related to a multi-view perspective \cite{glassner2000}. Not only cubists but also a wide variety of art schools used the multiview perspective for the creation of expressive depictions \cite{hess2004willem,hockney1995david}. All of the multi-view perspective techniques can be obtained with this approach. These include the creation of cameras using parametric surfaces \cite{glassner2000,smith2004,yu2004framework}, using images to control camera parameters \cite{meadows2000a,meadows2000b,morrison2020remote}, and using high-saliency regions to obtain cubist images \cite{collomosse2003cubist,arpa2012perceptual,wang2011cubist}. See Figure~\ref{fig_NPR} for examples. 

\section{Methodology}
\label{sec_methodology} 

\begin{figure*}[ht]
\begin{center}
\begin{tabular}{cccccc}
\includegraphics[width=0.32\textwidth]{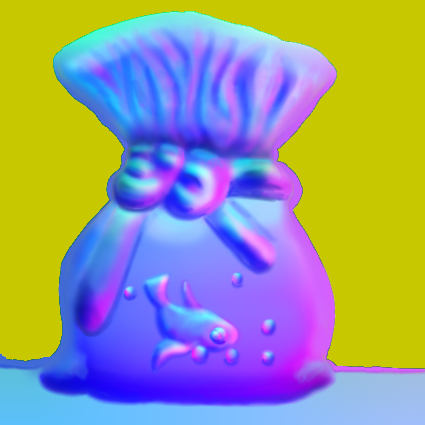}&
\includegraphics[width=0.32\textwidth]{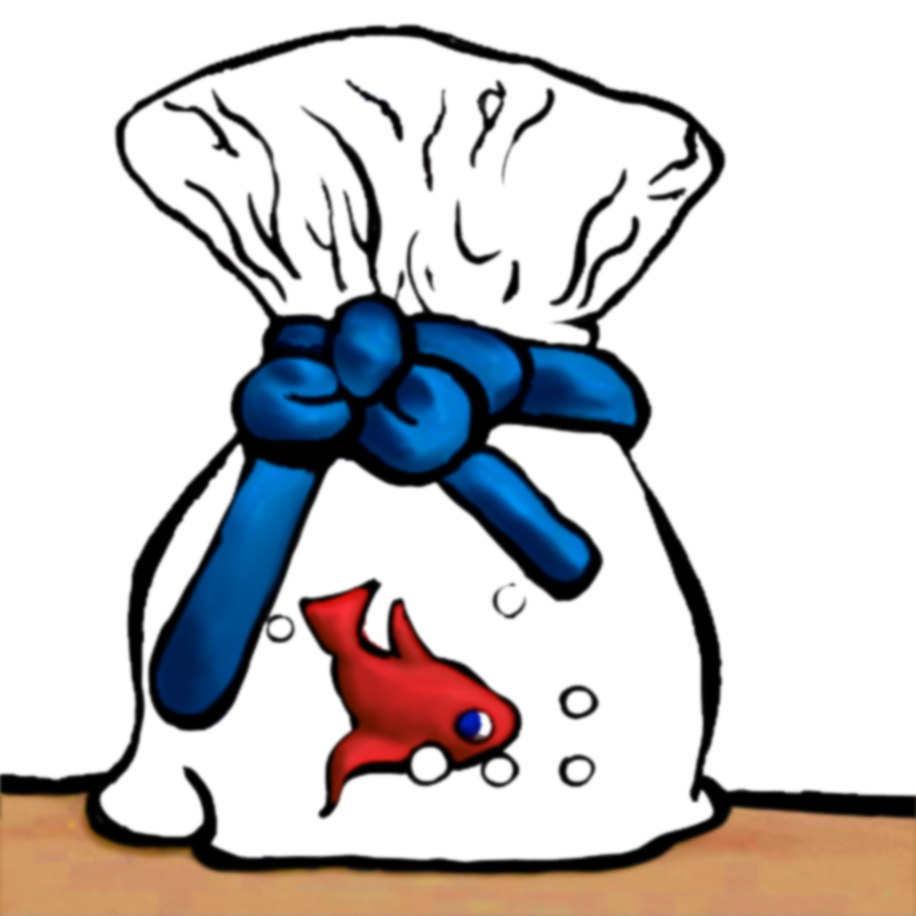}\\
(a) Shape Map &   (b) Foreground Image    \\
\end{tabular}
\begin{tabular}{cccccc}
\includegraphics[width=0.32\textwidth]{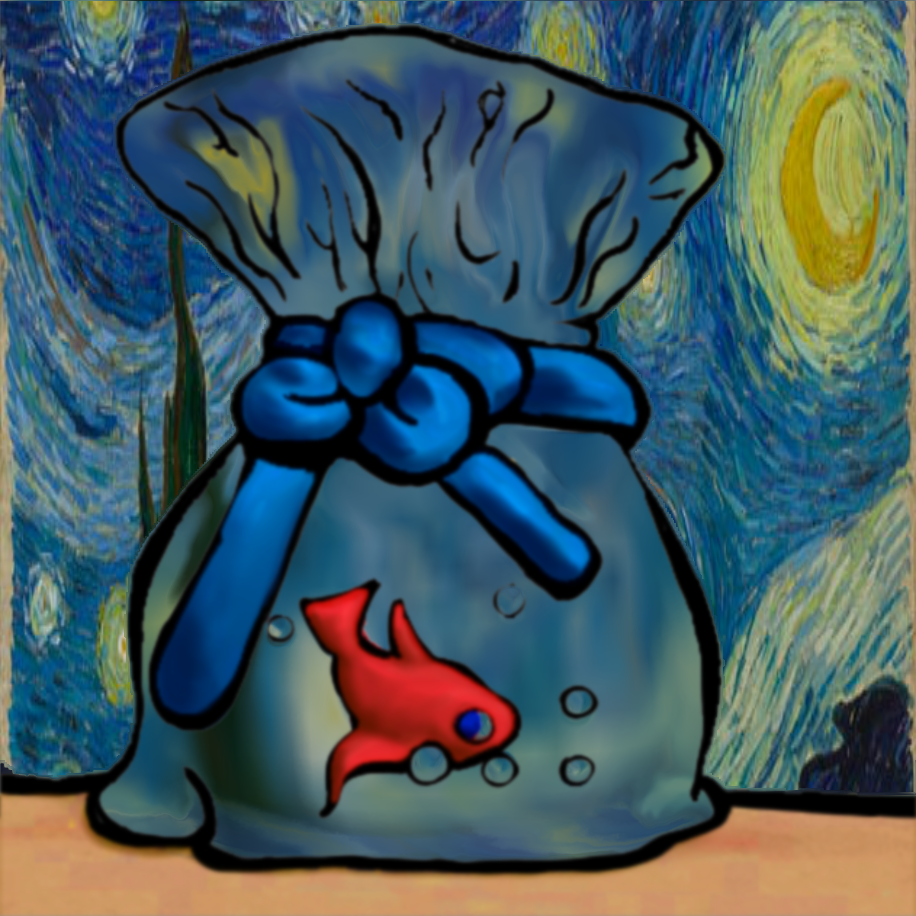}&
\includegraphics[width=0.32\textwidth]{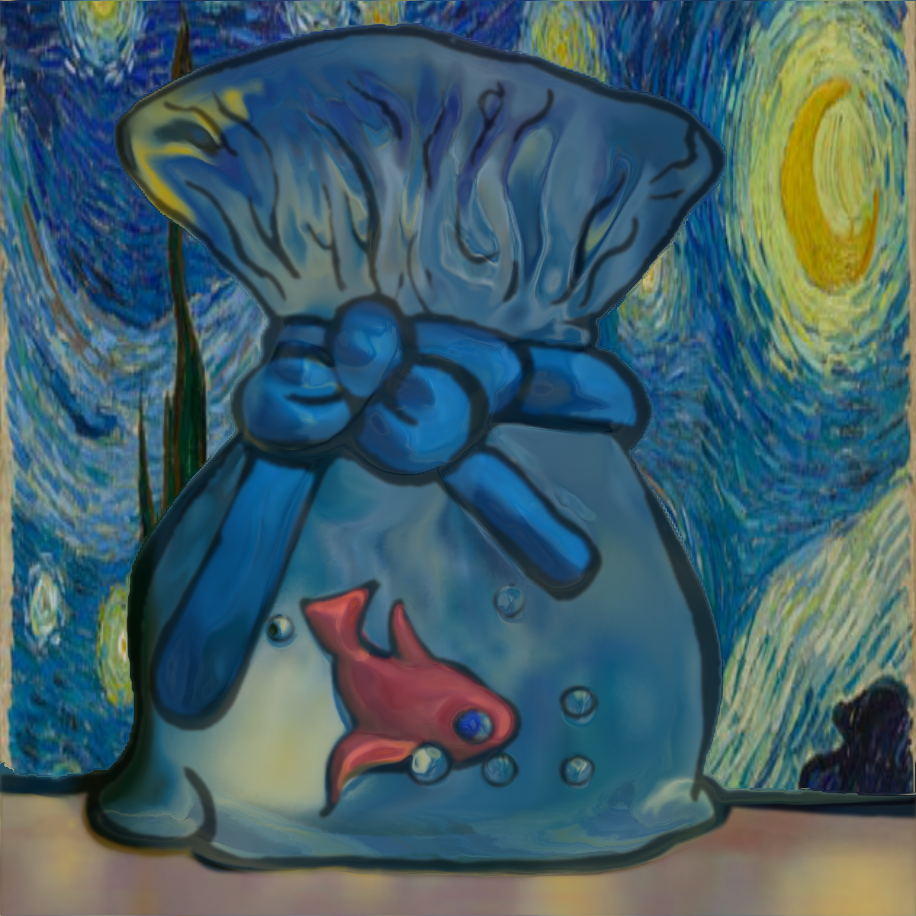}&
\includegraphics[width=0.32\textwidth]{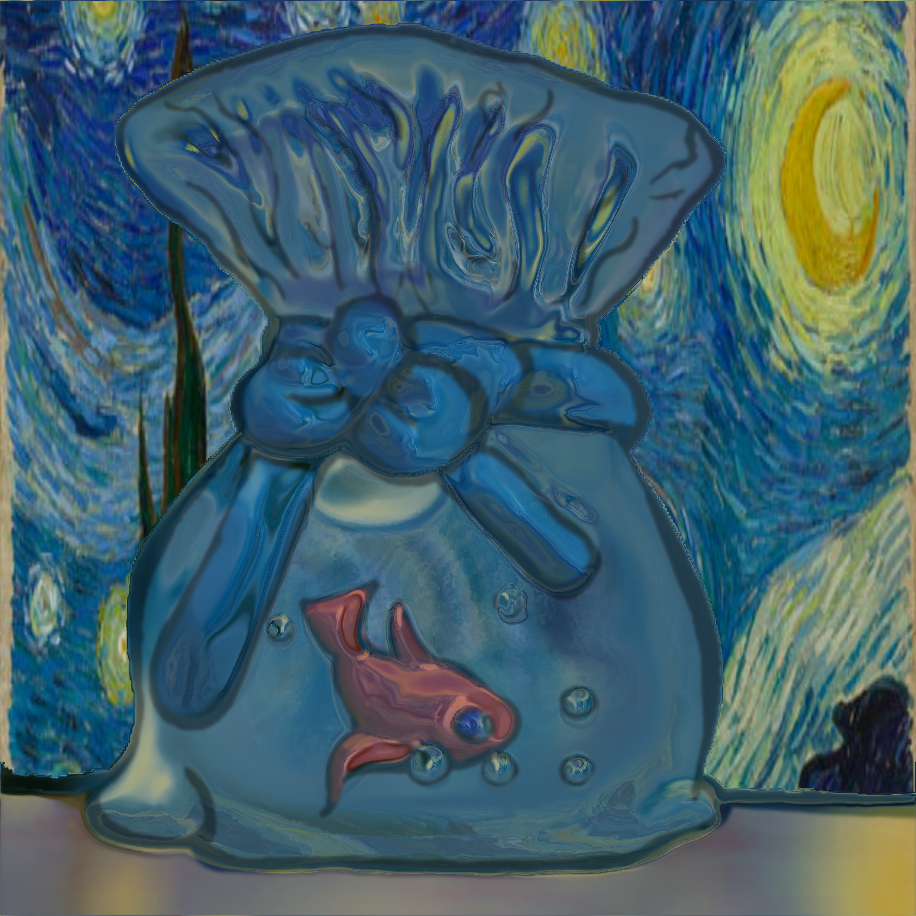}\\
   (c) Global Alpha $\alpha_G=1.$ &    (d)  Global Alpha $\alpha_G=0.5$  & (e)  Global Alpha $\alpha_G=0.25$ \\
\end{tabular}
\end{center}
\caption{\it An example of the effect of global $alpha$ with non-photo-realistic compositing with reflection, glossy reflection, refraction, and translucence combined with Fresnel. (a) Hand-drawn shape map and (b) foreground image. The white regions of the foreground image are transparent, i.e. $\alpha_{FI}=0$. In other regions, $\alpha_{FI}=1$. The images in (c), (d), and (e) show the effect of global $\alpha_G$. }
\label{fig_FishBall}
\end{figure*}

Our solution kills two birds with one stone by reformulating the compositing equation: (1) Our reformulation eliminates the need for 3D rendering to obtain realistic refraction and refraction effects, and (2) with our reformulation only one additional layer, namely shape map, is sufficient to obtain the composited image. In other words, a shape map combines all the mask and deformation information into one image. The new compositing equation that uses the shape map is given as follows:
$$CI = \alpha  FI + (1-\alpha) ( f R(EI) + (1-f) T(BI)).$$
where  $CI, FI, EI,$ and $BI$ are composite images, foreground images, environment maps, and background images, respectively. $R$ and $T$ are transformations that qualitatively provide reflection and refraction, respectively, and $f$ is the Fresnel term. $R$, $T$ and $f$ are calculated using a shape map, $SM$.  A rectangular environment map is used for reflection and a background image is used for refraction. The term $\alpha$ is computed as a multiplication of a global opacity term $\alpha_G$ with the alpha of the foreground image $\alpha_{FI}$. The effect of global alpha $\alpha_G$ is shown in Figure~\ref{fig_FishBall}.

\begin{figure*}[ht]
\centering
\begin{tabular}{cccccc}
\includegraphics[width=0.490\textwidth]{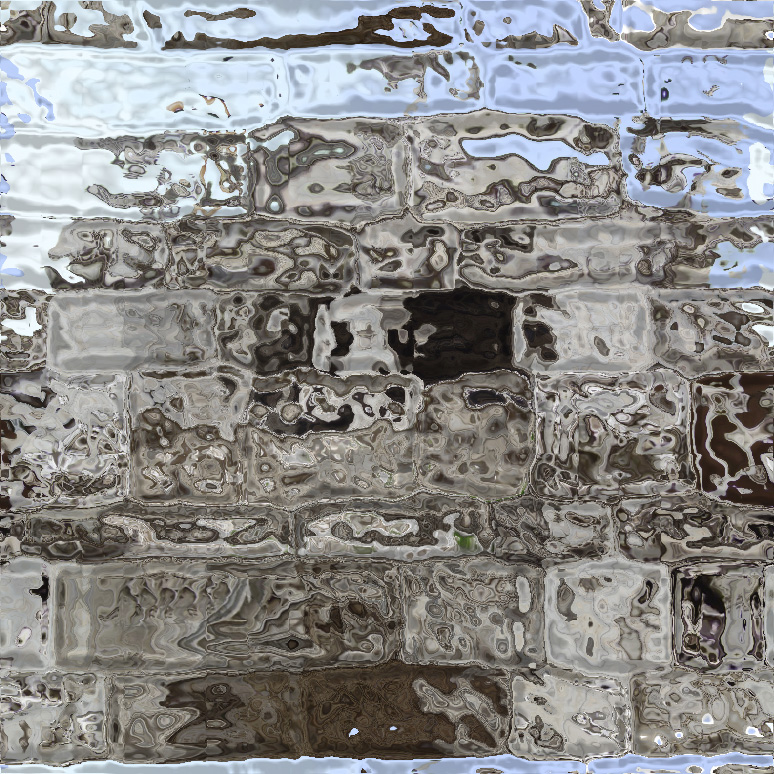}&
\includegraphics[width=0.490\textwidth]{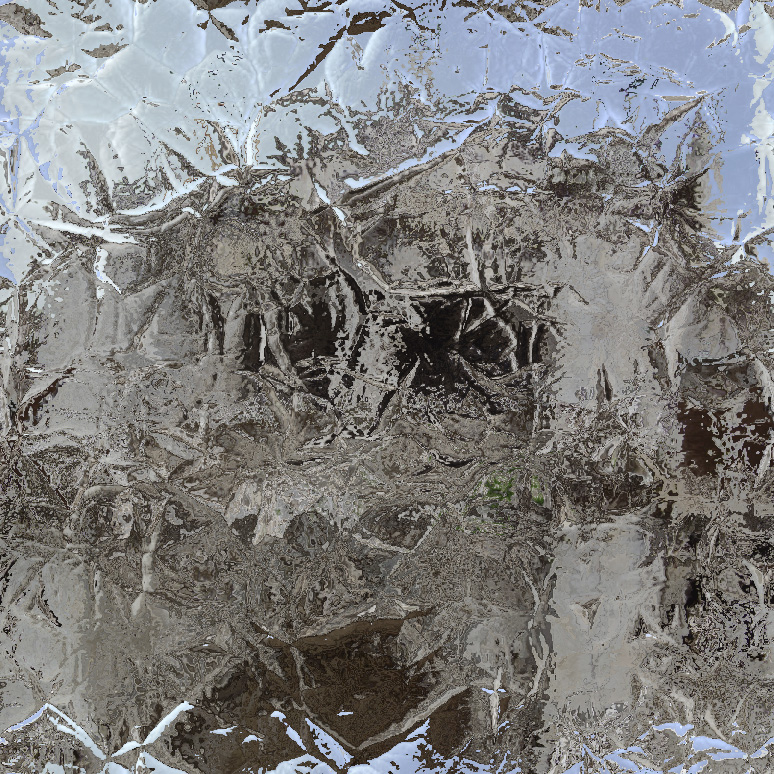}\\
\end{tabular}
\begin{tabular}{cccccc}
\includegraphics[width=0.320\textwidth]{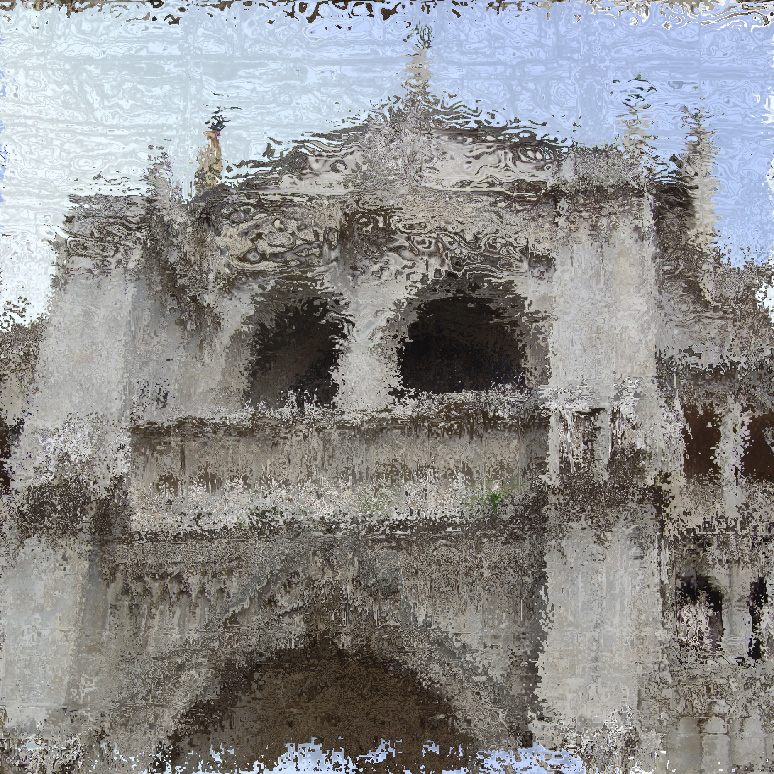}&
\includegraphics[width=0.320\textwidth]{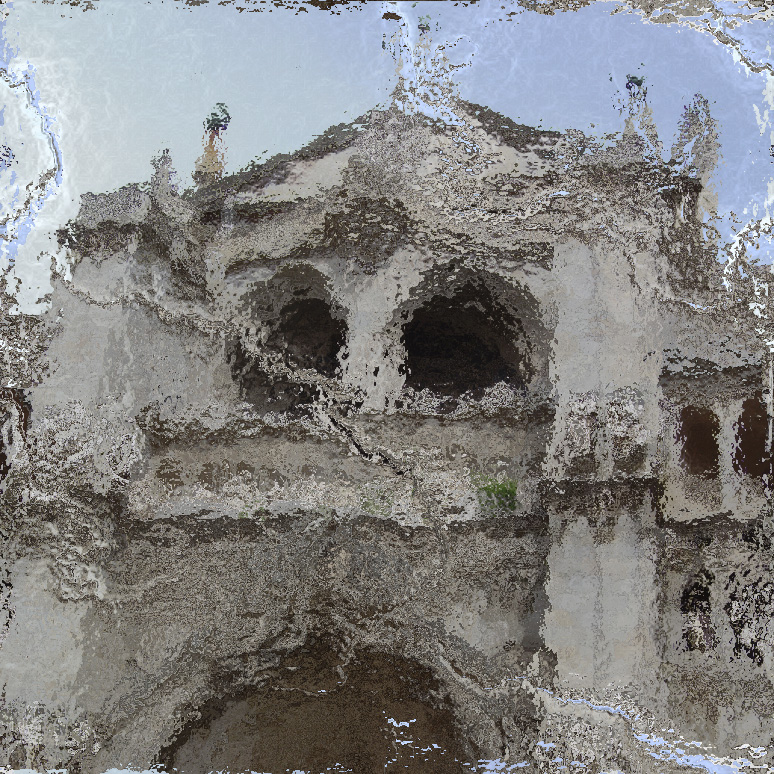}&
\includegraphics[width=0.320\textwidth]{NPR/result1}\\
\end{tabular}
\caption{\it Examples of painterly filter effect obtained by our compositing equation using with variety of shape maps. In these examples, we use only a background image and a shape map. One can create more complicated images by including a reflection from the environment map and a foreground image. Using this approach, it is also possible to obtain \cite{akleman2023recursive}.}
\label{fig_NPR}
\end{figure*}

The major advantage of this formulation is that any image can be used as a shape map. In other words, shape maps do not have to correspond to any real shape. Therefore, artists can create shape maps to obtain purposely non-realistic images such as cubist paintings, multi-perspective images, expressionist, and abstract paintings. Since shape maps are like shower doors, one can directly control painterly effects by painting shape maps as shown in Figure~\ref{fig_NPR}. These images can also be dynamically manipulated by changing the refraction properties.

\begin{figure}[ht]
\begin{center}
\begin{tabular}{cccccc}
\includegraphics[width=0.320\textwidth]{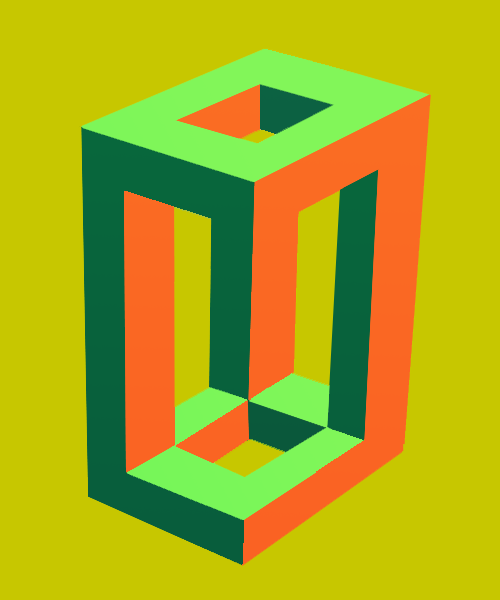}&
\includegraphics[width=0.320\textwidth]{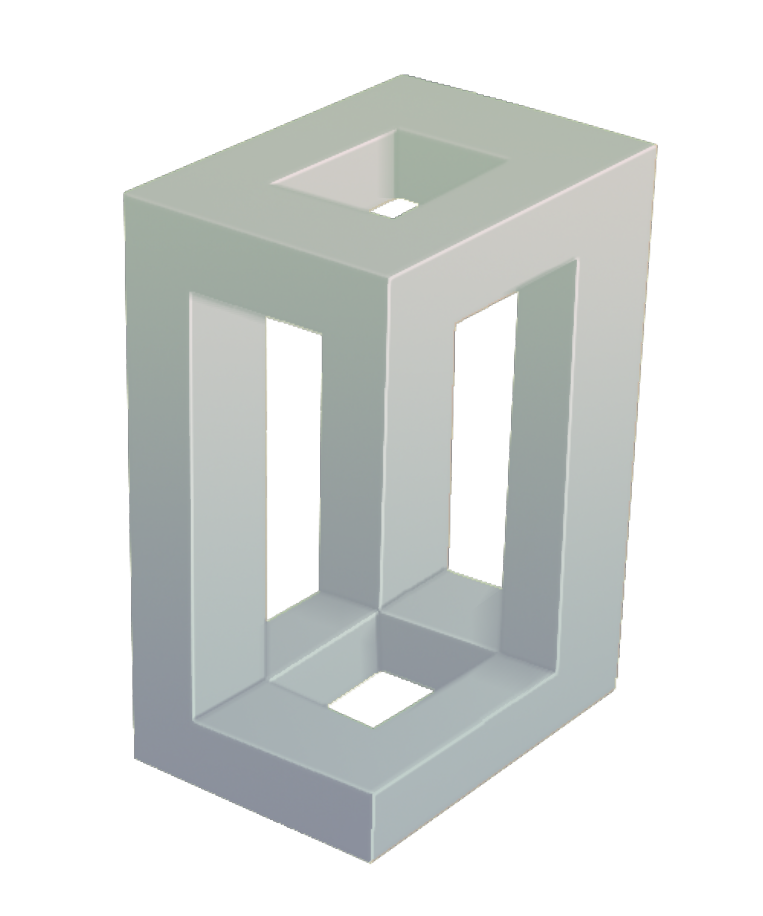}&
\includegraphics[width=0.320\textwidth]{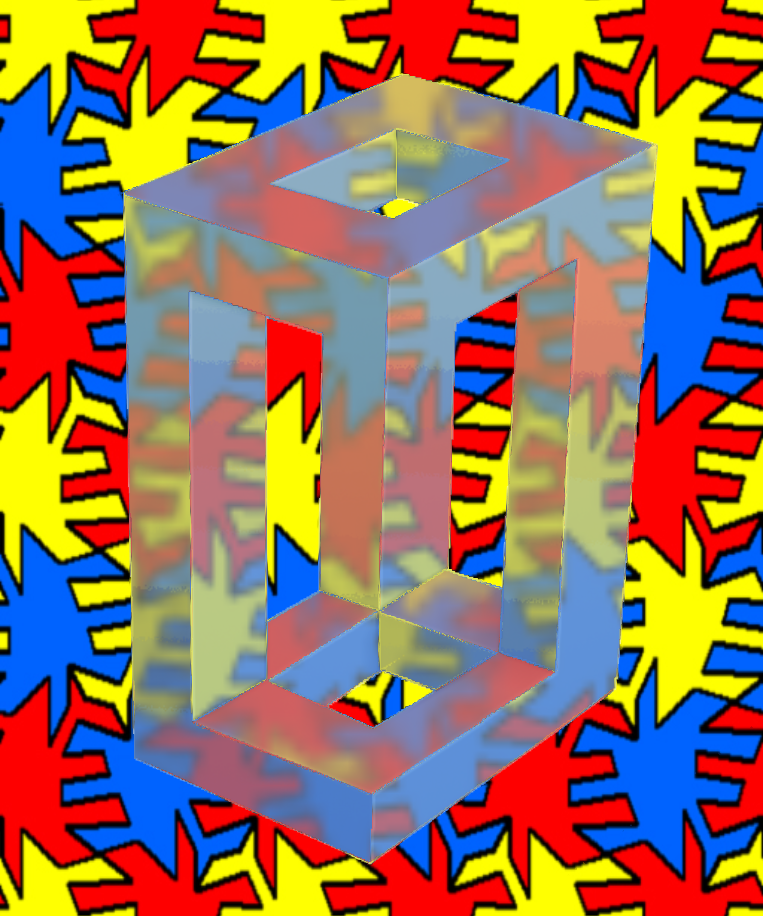}\\
Shape Map &    Foreground &    Compositing  \\
\end{tabular}
\end{center}
\caption{\it An example of reflection and refraction with an impossible object. The shape map and the foreground image were painted by an artist in a digital painting program. We use the global $\alpha_G=0.5$ to make the foreground layer slightly transparent. Note that there exists no continuous height field that can produce these vector fields.}
\label{fig_impossible}
\end{figure}

Using the same approach, it is also possible to obtain reflection and refraction with impossible shapes (see Figure~\ref{fig_impossible}). For some impossible shapes, it is possible to construct 3D shapes that look impossible from one point of view \cite{Elber2011}, however, even they cannot be made transparent without revealing that they are not impossible. The images in Figure~\ref{fig_impossible} are the first examples of impossible transparent objects to our knowledge. 

It should be noted that Ritschel et al. are the first to explore reflection/refraction in art painting \cite{ritschel2009approximating}. Although they provide solutions departing from the laws of physics,  their solutions are still in 3D and require the specification of a 3D scene such as objects, lights, and materials. Therefore, their solution allows only modification of existing effects but does not provide complete control of the final results. Our solution is based on reformulating the compositing equation using an extended version of normal maps, providing a truly 2D solution with complete user control of the results. Our approach is more related to surface flows, which are introduced by Vergne et al. \cite{Vergne2012} to interactively manipulate normal maps to obtain desired shading, reflection, and refraction effects.

Shape maps are essentially an extension of normal maps \cite{Cohen1998}. Although normal maps are usually used as texture maps to include details of polygonal meshes, they can also be used as shape representations. Johnston developed Lumo to model normal maps as mock-3D shapes by diffusing 2D normals in a line drawing \cite{Johnston2002}. Since then, several groups have investigated modeling normal maps as mock-3D shape representations \cite{Okabe2006,Shao2012}. To obtain refraction, the normal map is useful, but the thickness of the shape needs to be known to explicitly control the refraction effects. Shape maps are essentially normal maps with an additional displacement parameter that corresponds to the thickness of the shape.

Let a square $M= [0,1] \times [0,1]$ denote a shape map and let $(u,v) \in M$ denote two coordinates of the map. Let $(x(u,v),y(u,v))$ denote the 2D vector field and the displacement (or thickness) parameter $d(u,v)$ with $x : M \rightarrow [-1,1], $, $y : M \rightarrow [-1,1], $ and  $d : M \rightarrow [0,1].$\\
The $x,y$ components of the 2D vector field can be both negative or positive, but the displacement parameter is always positive. The new vector field $(dx,dy)$ is used to realistically calculate the deformation caused by refraction with an artistic control. The parameter $d$ alone is also used to define a mask, i.e. the object exists only in the regions where $d \neq 0$. The displacement parameter $d$ is qualitatively related to the thickness of the transparent object. 

\begin{figure}[htpb]
  \centering  
\begin{subfigure}[t]{0.432\textwidth}
  \fbox{\includegraphics[width=1.0\textwidth]{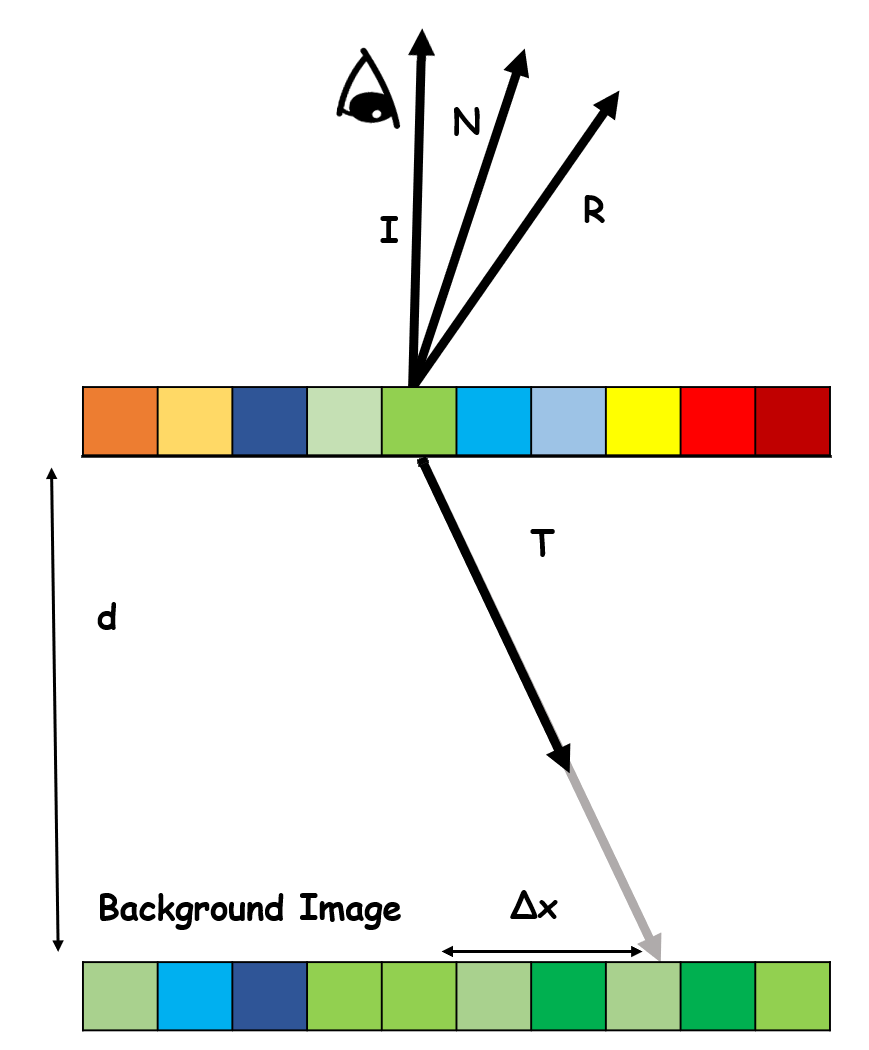}}
  \caption{
  2.5D Refraction with Shape Maps. Side view.}
  \label{figimages/R1}
  \end{subfigure}
  \hfill  
  \begin{subfigure}[t]{0.535\textwidth}
  \fbox{\includegraphics[width=1.0\textwidth]{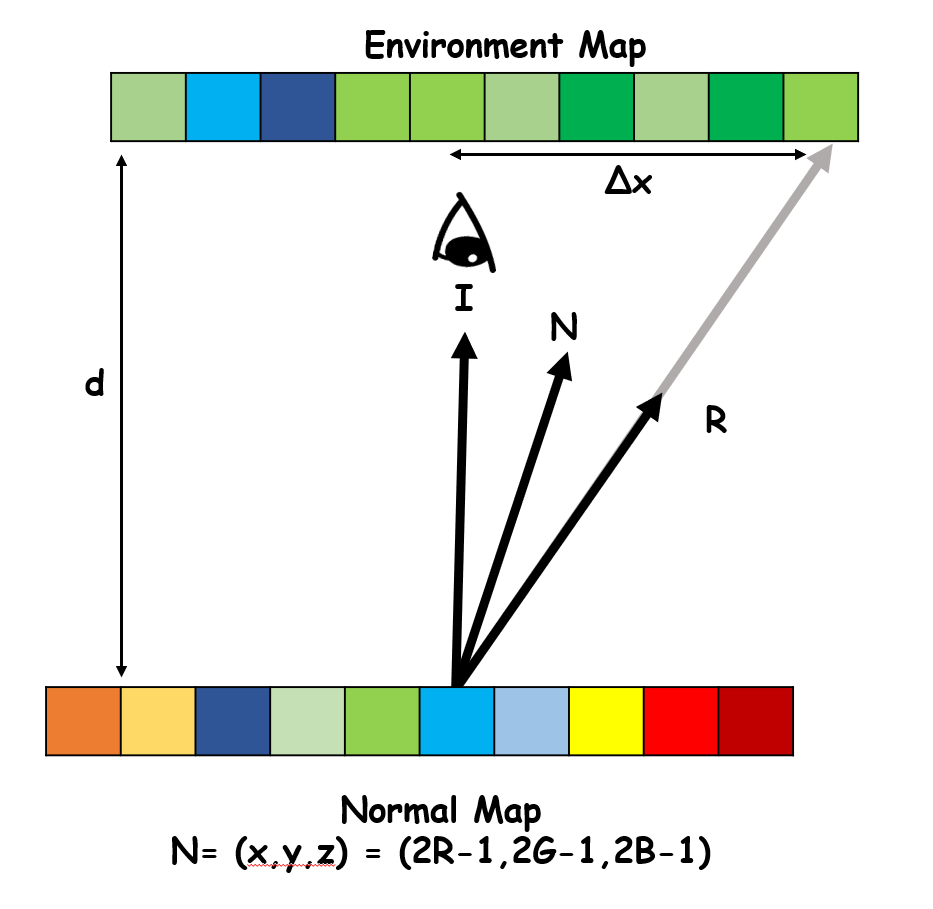}}
  \caption{
  2.5D reflections with Shape Maps. Side view.}
  \label{figimages/R0}
  \end{subfigure}
  \hfill  
\caption{Computation of the reflection and refraction in 2.5D space using Shape Maps. Here, $\vec{I}$ is the eye vector, $\vec{N}$ is the normal vector, $\vec{R}$ is the reflection vector, and $\vec{T}$ is the transmission vector. For refraction, the term $d$ is given by the shape map and gives the distance between the shape map and the background image (see Figure~\ref{figimages/R1}). The other term $d$, which is the distance between the Shape Map and the Environment Map, is used to calculate the reflection (see Figure~\ref{figimages/R0}). This particular term is a global parameter and is constant for all pixels. See \cite{akleman2023webbased} for the algorithm.  }
\label{figimages/R}
\end{figure}

We observe that thickness is more important than the index of refraction in controlling the deformations caused by refraction. For instance, a thin glass will hardly cause any deformation, but a thick volume of water will. For most objects, the thickness is not constant, and this variance in thickness provides the kind of visual effects that we humans are familiar with. For example, to create realistic effects, it is important to choose small values of $d$ around the 2D boundaries of the objects. However, the actual values of  $d$  in other parts of the objects are not that critical. Figure~\ref{figimages/R1} shows the impact of the term $d$. 

If we consider $x,y$ of normal maps as a 2D vector field and $z$ as displacement, any normal map can be used directly as a shape map. This is not an unexpected result. If the 3D normal vector is a unit vector, a 2D vector field, $(x,y)$, is sufficient to extract the third-dimensional information, since the value of $z$ can always be calculated as $z(u,v) = \sqrt{1-x^2-y^2}$ ($z$ can never be negative). This assumption is correct only if normal vectors are computed correctly as unit vectors from a given 3D shape. Fortunately, even when the 2D vector field is imprecise, it is still possible to assume that the $z$ component is implicitly provided. The main issue is that in shape maps it is possible to have $x^2+y^2>1$ since the shape map data can be created in an imprecise way by users. This is not a problem, since we can always assume that the 2D vector field is uniformly scaled by a parameter $w \in [0,1]$. For values $w$ smaller than $1/\sqrt{2}$, the unit 3D normal vector always exists $n(u,v) =( wx, wy, \sqrt{1-w^2 x^2-w^2 y^2})$.

We want to point out that scaling in the 2D vector domain corresponds to flattening of the 3D shape and is a visual equivalent of flattening sculptures into bas reliefs \cite{Weyrich2007}. Therefore, the rendering results are not expected to create visual problems, and visually acceptable results are possible from imperfectly defined 2D vector fields using low values $w$. For high-quality shape maps, it is always possible to use $w=1$. For others, a value smaller than $w=0.7$ always works. In practice, our default value for $w$ is $w=0.5$.

Another commonality with normal maps is that the value $z$ in normal maps can be used as the displacement parameter $d$.  for the refraction of the boundaries. Setting displacement values around silhouette edges is crucial to obtaining realistic-looking refraction.  Since the value of $z$ is naturally low around the silhouette edges of the shape, we can directly use the values of $z$ as $d$ at the limits. However, the similarity ends with boundaries. In cases such as the fish bowl and wine bottle (see Figures~\ref{fig_FishBall} and~\ref{fig_Bottle}) the values of $d$ must be set independently of $z$ to obtain a realistic refraction.

The advantage of shape maps is that they are not required to correspond to any given shape. Even when there is no imperfection in the data,  there may still not exist any height field whose gradient can produce the given 2D normal vector field. To obtain a height field that can minimize errors, we have to solve the Poisson equation \cite{Fattal2002}. This expensive computation is unnecessary for shape maps.  Shape maps can be easily obtained using 3D rendering software. The procedure for obtaining a 2D vector field is a straightforward rendering process. The $x$ and $y$ components of the 3D normal vector of the visible point are simply converted to red and green in the image. As discussed above, it is also possible to use the component $z$ of the unit normal vector as the value of $d$. 

This choice will provide small $d$ values on the boundaries of the object. However, the object may be thin even if the value of $z$ is not small. One such example is the shape map of the wine bottle shown in Figure~\ref{fig_Bottle}. This bottle is half full of liquid. In the places where the bottle is not filled with liquid, the value of $d$ must be very small, although the $z$ component of the unit vector is not small. For such cases, the values of $d$ in $(u,v)$ can be computed as the length of the line of $(ut, vt, t)$ that is within the corresponding 3D shape. The problem with this approach is that it requires separate 3D rendering software to compute shape maps. For artistic control, it is more important to design shape maps directly in 2D using painting or drawing software.

\section{Designing Shape Maps Directly in 2D \label{section:SMIP}}

One significant advantage of using only three variables for shape maps is that we can readily convert shape maps into low-dynamic range (LDR) images and save them using any common image format that can easily be passed to GPU. We assume that an LDR image on $M$ is denoted by $c(u,v) = (r(u,v),g(u,v),b(u,v))$ where $c: M \rightarrow [0,1]^3$. The conversion from $(x,y,d)$ to $(r,g,b)$ is given as $(r=0.5(x+1),g=0.5(y+1),b=d)$. As can be seen in the figures, the shape map images are more colorful than normal maps, since we use blue color as independent displacement information. The main advantage to thinking of shape maps as 3-color images is that artists can create shape maps directly using 2D painting software.

To paint shape map images, the painters do not need to think that they are working on a gradient domain. They can imagine an object lit by 2-point lighting illuminated from left with a directional (parallel) red light and from above with a directional green light. Ignoring shadows, they can paint the image based on how much red and green light they want to see in every pixel. For example, a pixel color red=0.75 and green=0.3 means that the artist wants 75\% of the light from the left and 30\% of the light from the top to illuminate that particular pixel.

\begin{figure}[ht]
\begin{center}
\begin{tabular}{cccccc}
\includegraphics[width=0.45\textwidth]{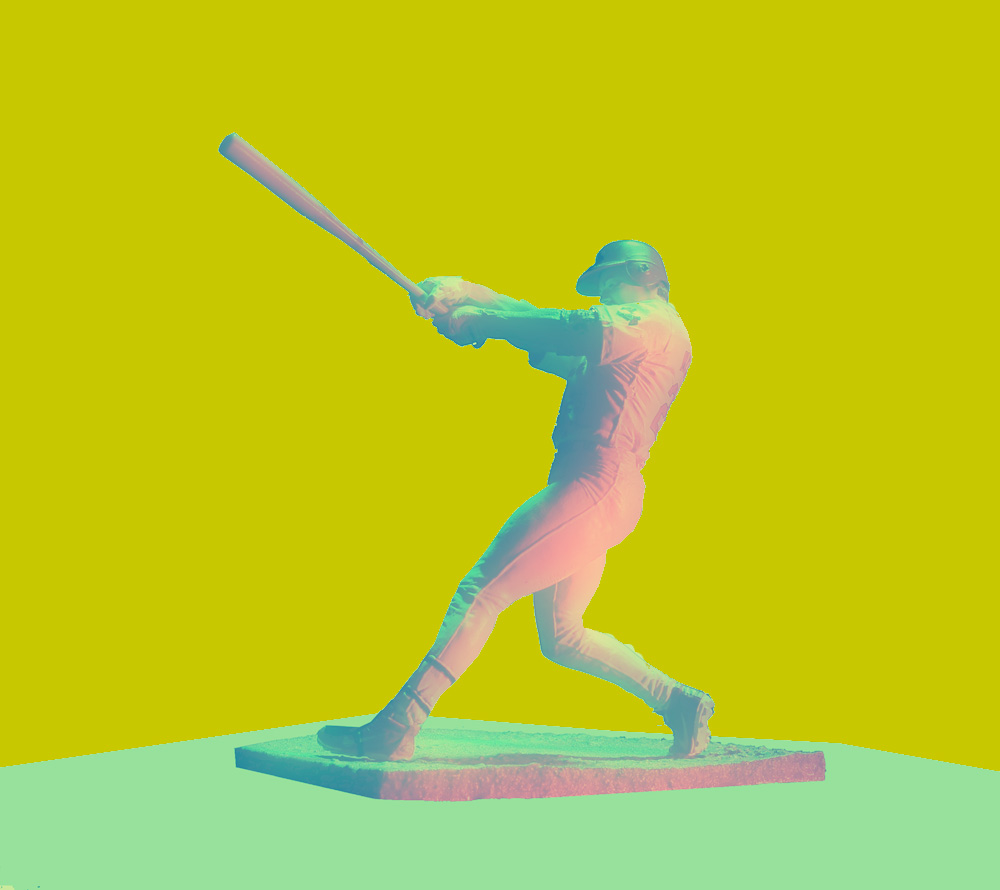}&
\includegraphics[width=0.45\textwidth]{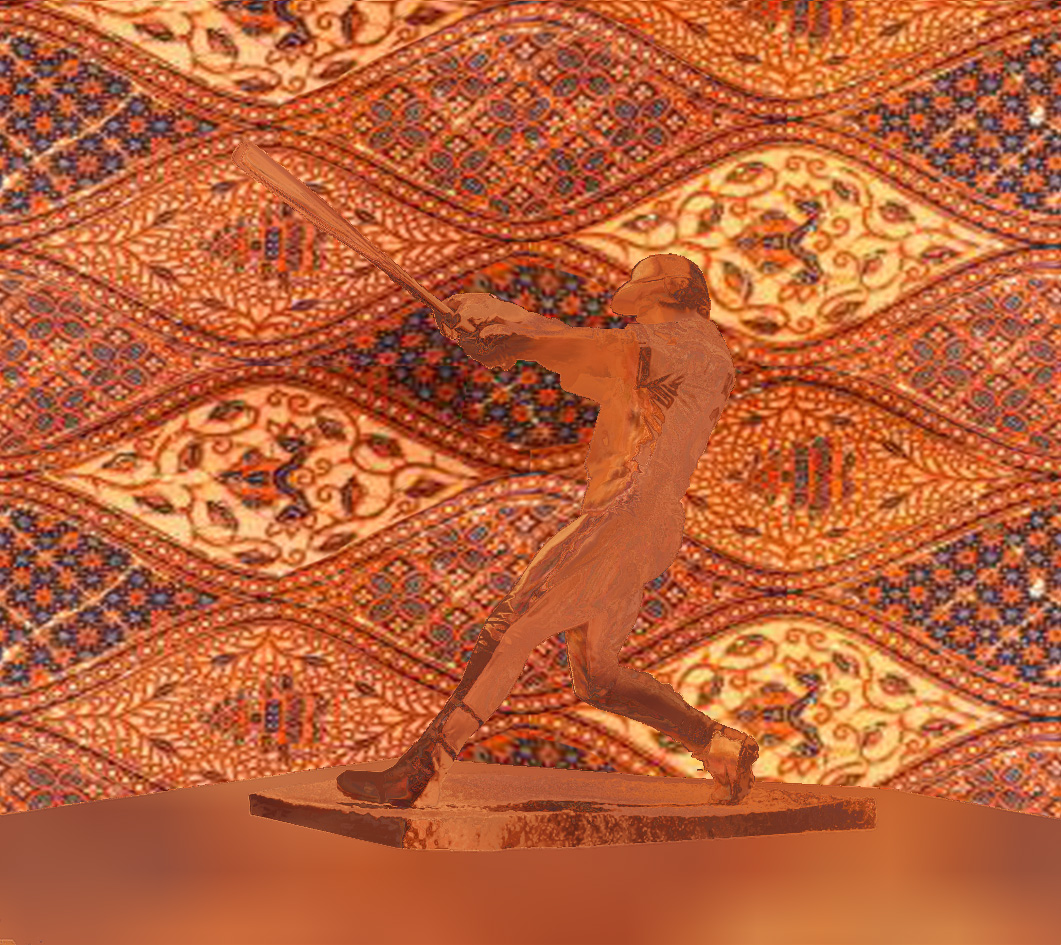}\\
\includegraphics[width=0.45\textwidth]{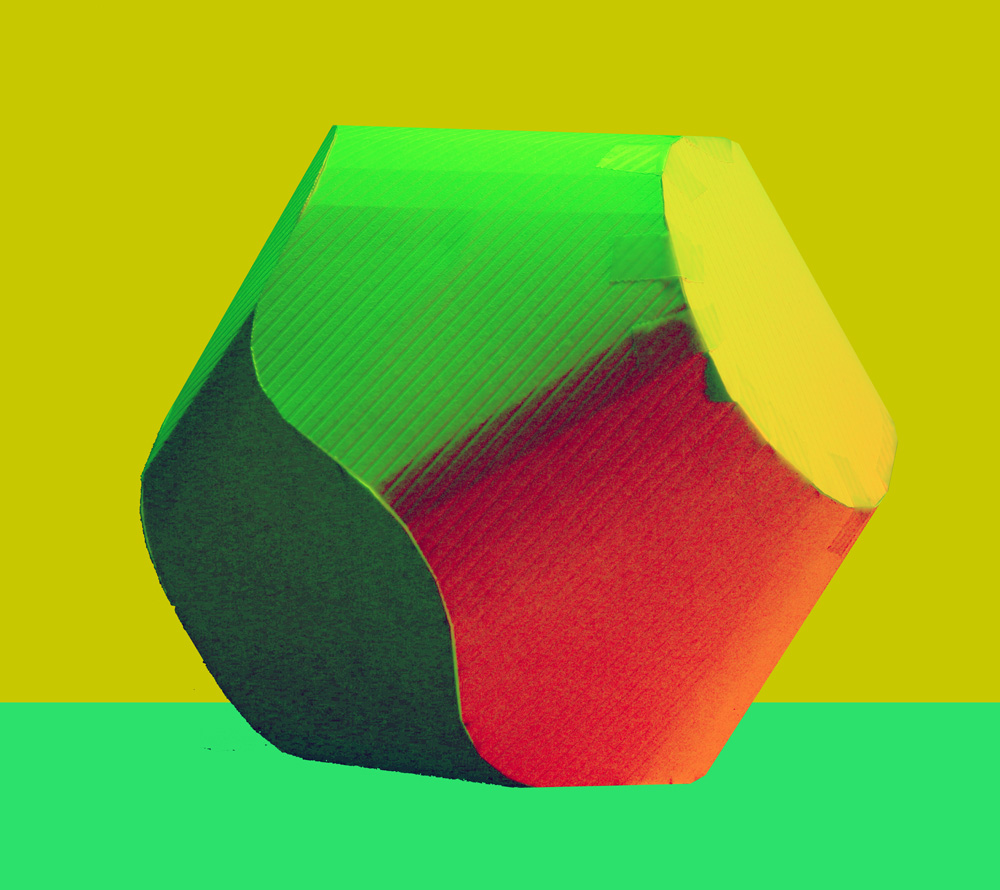}&
\includegraphics[width=0.45\textwidth]{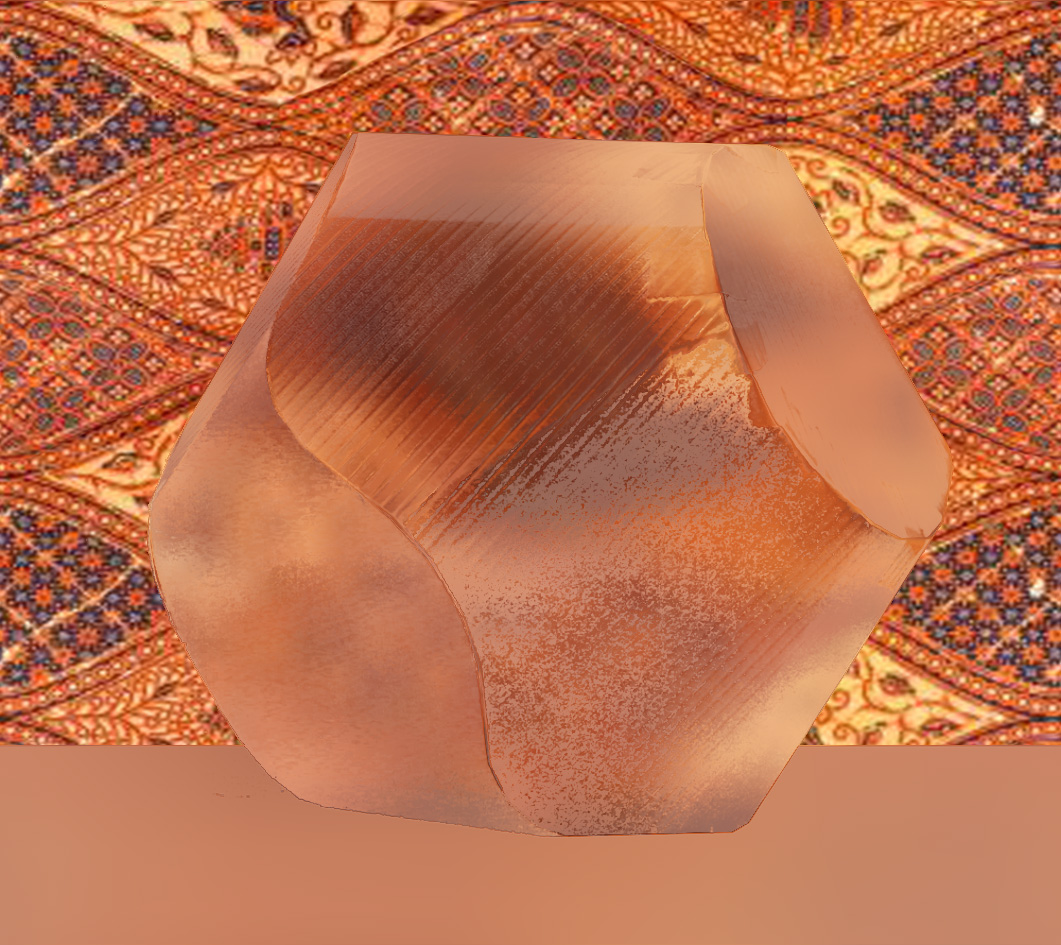}\\
Shape Map Photos &    Composite  \\
\end{tabular}
\end{center}
\caption{\it Examples of shape map photographs of diffuse objects. In these cases, we do not have any foreground object, that is, $\alpha=0$. Composite images are simply the result of refraction and refraction of the same image that is used for both the environment and the background image. The first object is just a figurine of a baseball player made of plastic. The second object is a D-form (a developable polyhedron) made of paper. }
\label{fig_photos0}
\end{figure}

These red-green paintings, although imperfect, provide good estimations of 2D gradient domains. Since the red and green light vectors $(1,0)$ and $(0,1)$ are linearly independent of each other, any 2D light can be given a linear combination of the two as $(x_L,y_L)=x_L(1,0)+ y_L(0,1)$. Therefore, to compute illumination coming from an arbitrary parallel light, we simply compute the contribution from two linearly independent components. 

\begin{figure}[htb!]
\begin{center}
\begin{tabular}{cccccc}
\includegraphics[width=0.45\textwidth]{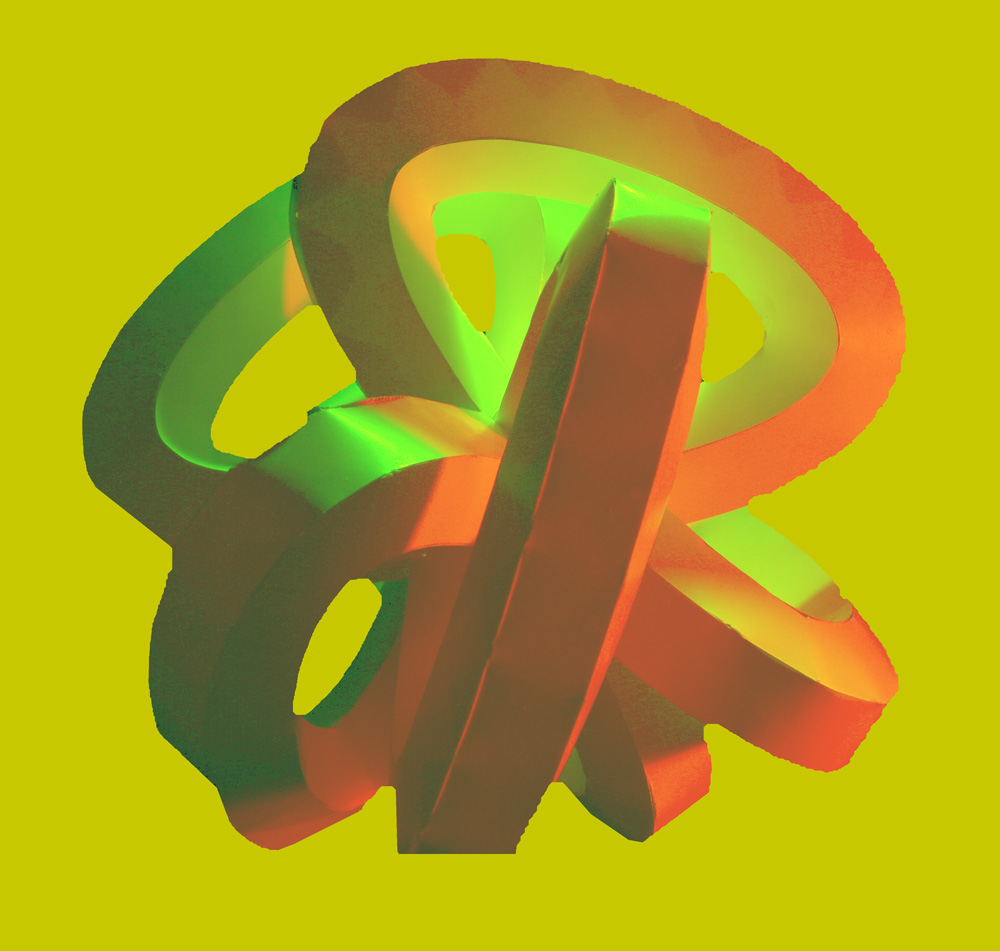}&
\includegraphics[width=0.45\textwidth]{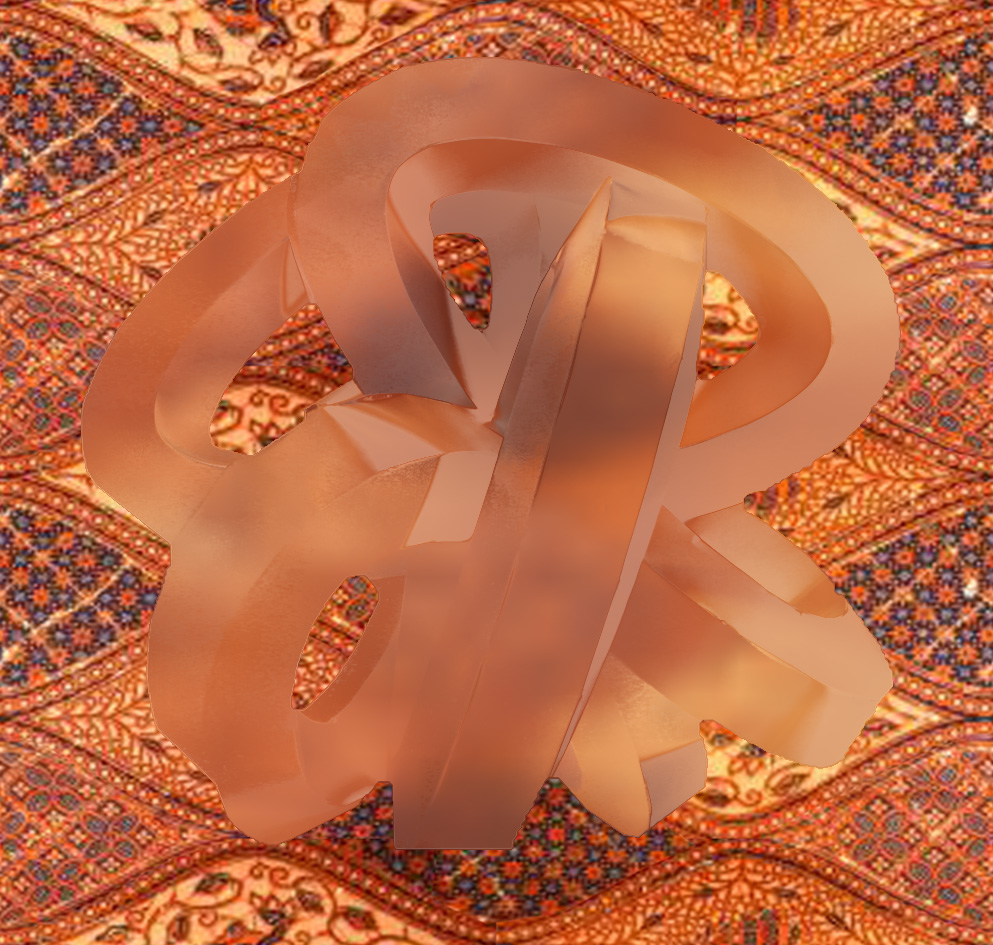}\\
\includegraphics[width=0.45\textwidth]{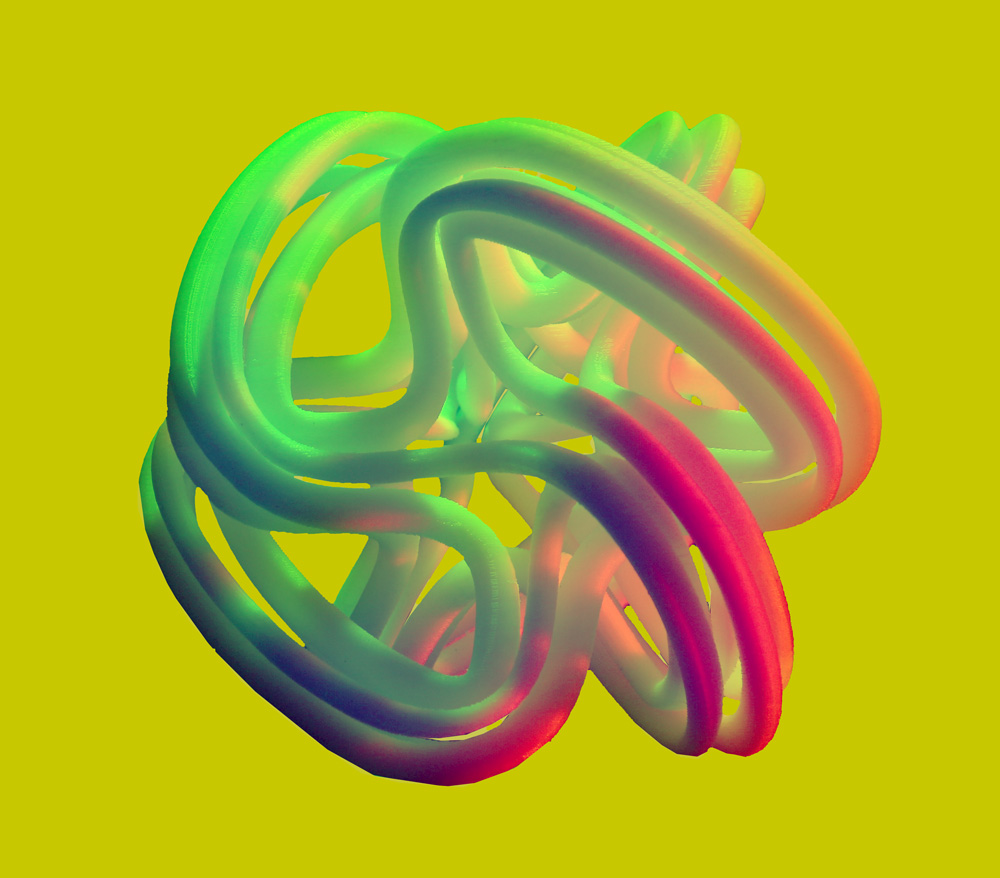}&
\includegraphics[width=0.45\textwidth]{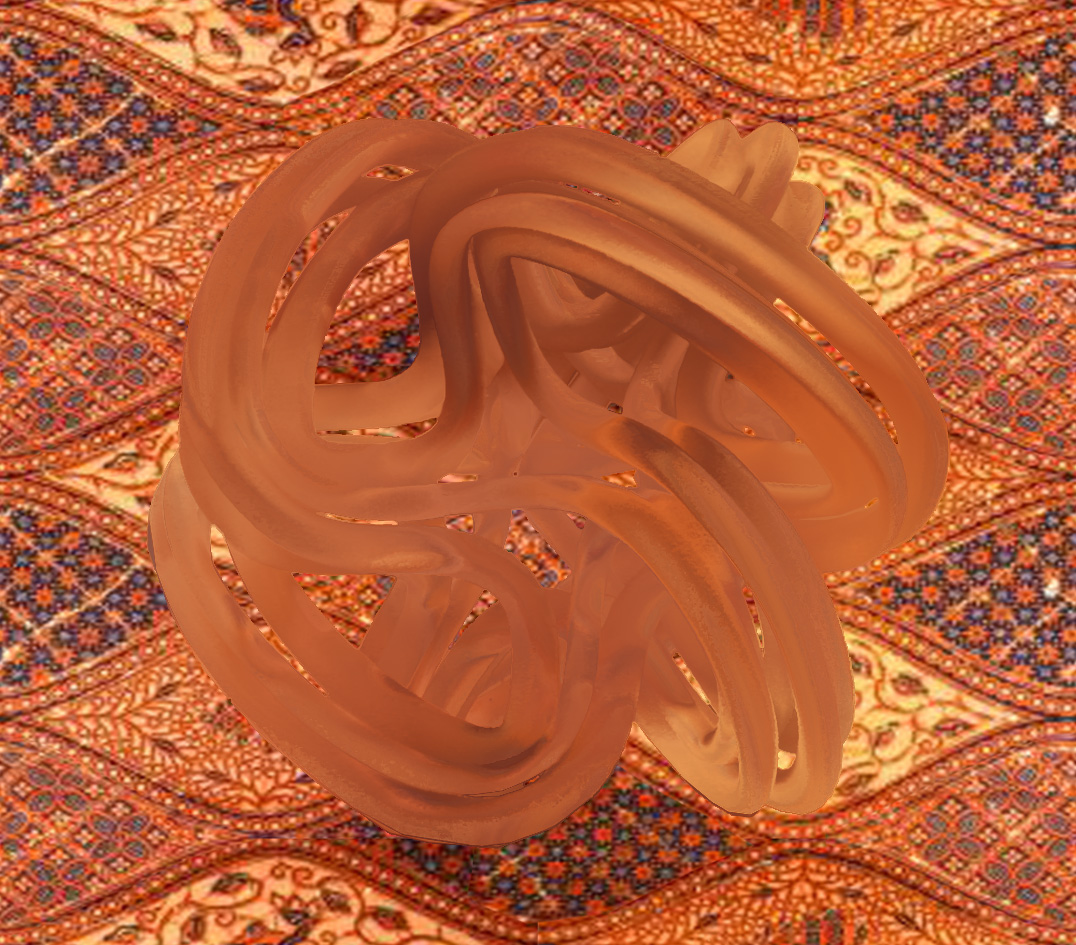}\\
Shape Map Photos &    Composite  \\
\end{tabular}
\end{center}
\caption{\it Examples of shape map photographs of diffuse objects. In these cases, we do not have any foreground object, that is, $\alpha=0$. Composite images are simply the result of refraction and refraction of the same image that is used for both the environment and the background image. The original of the object of genus 6 at the top is made of paper. The object of the high genus in the middle is made of ABS plastic}
\label{fig_photos1}
\end{figure}

Painting the values of $d$ is easier. As discussed earlier, the $d$ values must be non-zero for the object and zero for everywhere else. Moreover, the values of $d$ must be smaller close to the boundary of the object and in thin regions. This approximation is sufficient to obtain visually correct-looking refractions. For the rest of the object, the values $d$ can simply be any reasonable positive real number less than $1$. Artists can also experiment with $d$ values to obtain unusual effects.

Providing direct control to artists is essential for the creation of unusual images, since artists can deliberately introduce imperfections that lead to expressive and artistic effects. For example, the imperfections introduced by the artist on the shape map of the wine bottle shown in Figure~\ref{fig_Bottle} resulted in Fresnel calculations cross-hatching when the background was completely black and the reflected environment was white. Other examples of shape maps painted by artists are shown in Figures~\ref{fig_FishBall} and~\ref{fig_impossible}.

\begin{figure}[htb!]
\begin{center}
\begin{tabular}{cccccc}
\includegraphics[width=0.45\textwidth]{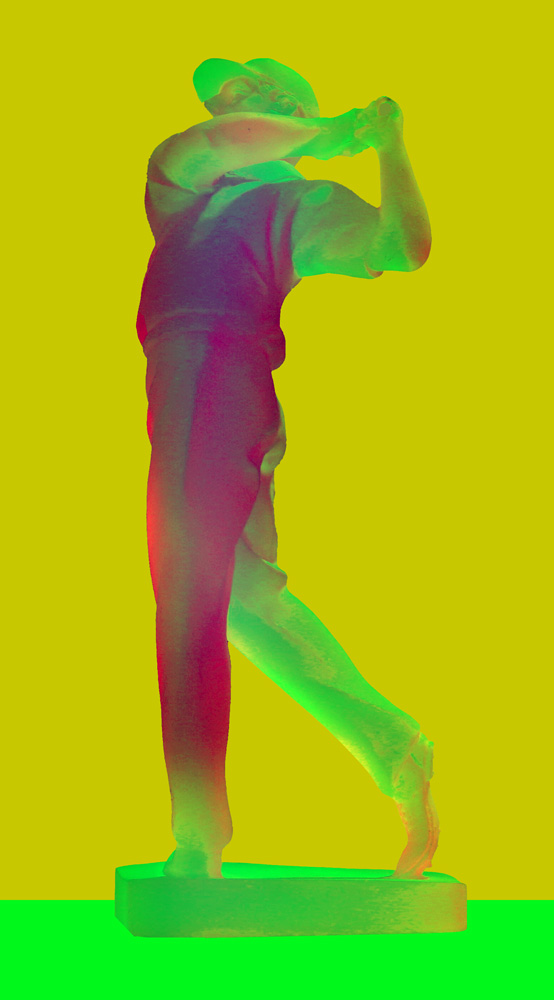}&
\includegraphics[width=0.45\textwidth]{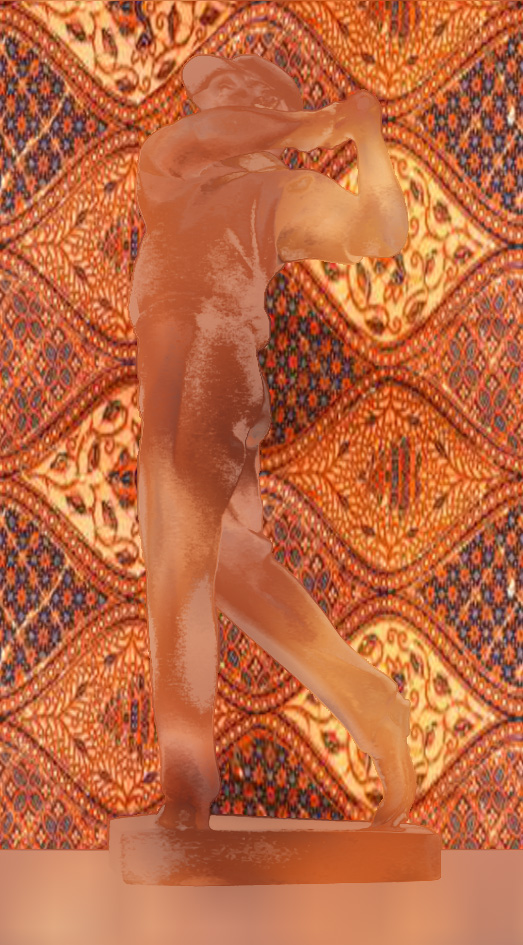}\\
Shape Map Photo &    Composite  \\
\end{tabular}
\end{center}
\caption{\it An example of a shape map photograph of a translucent object. In this case, we do not have any foreground object, that is, $\alpha=0$. Composite images are simply the result of refraction and refraction of the same image that is used both for the environment and for the background image.}
\label{fig_photos2}
\end{figure}

\begin{figure}htb!]
\begin{center}
\begin{tabular}{cccccc}
\includegraphics[width=0.45\textwidth]{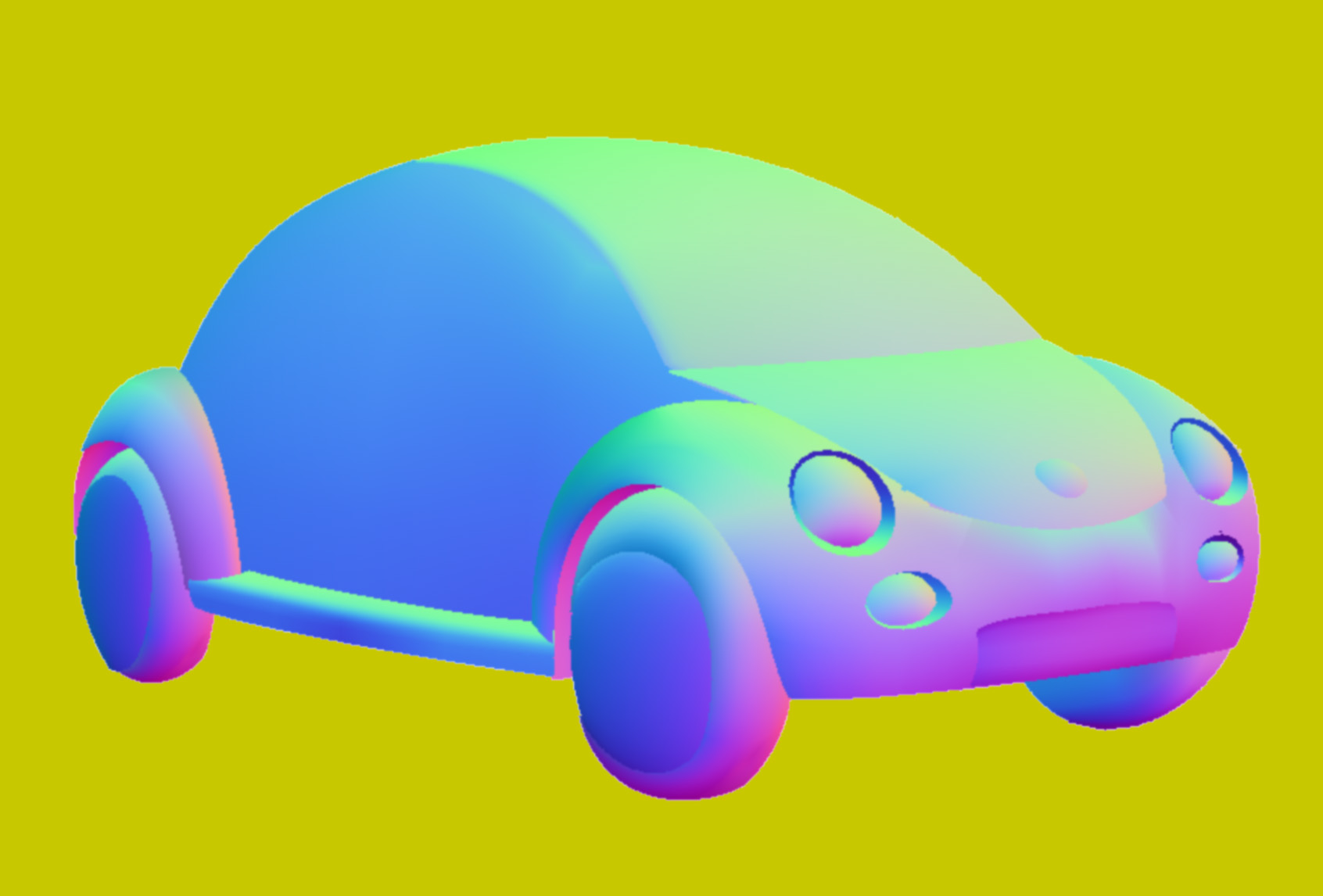}&
\includegraphics[width=0.45\textwidth]{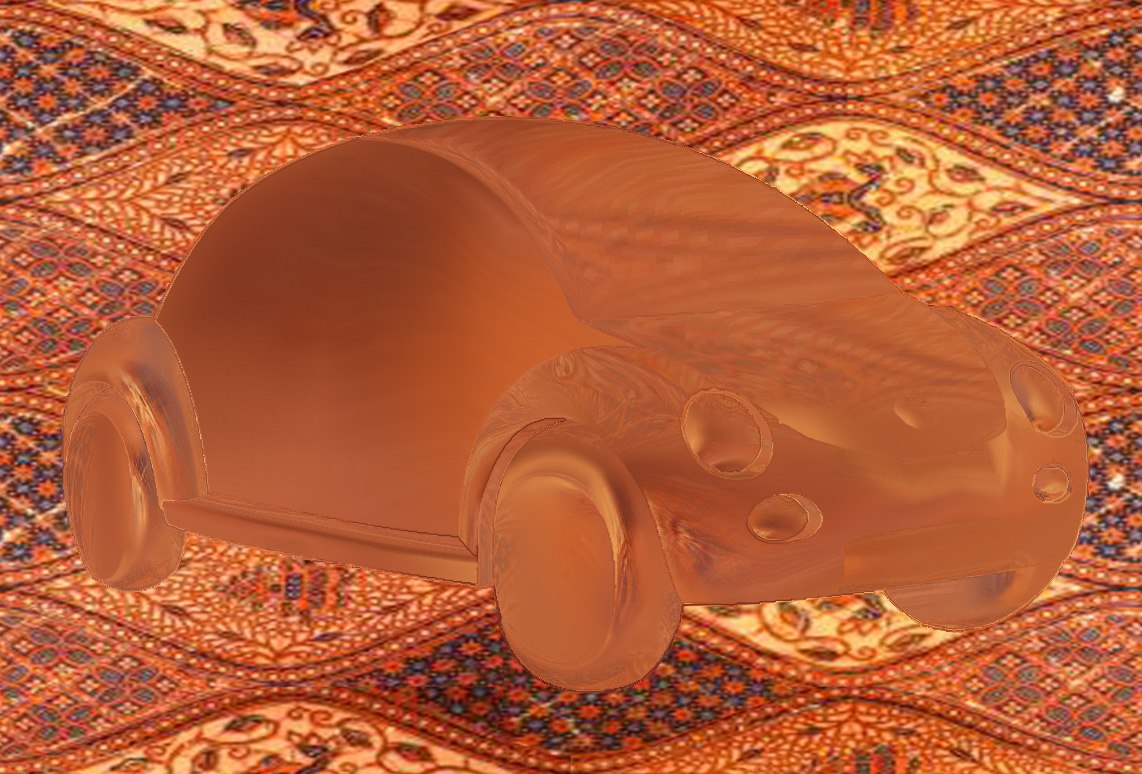}\\
\includegraphics[width=0.45\textwidth]{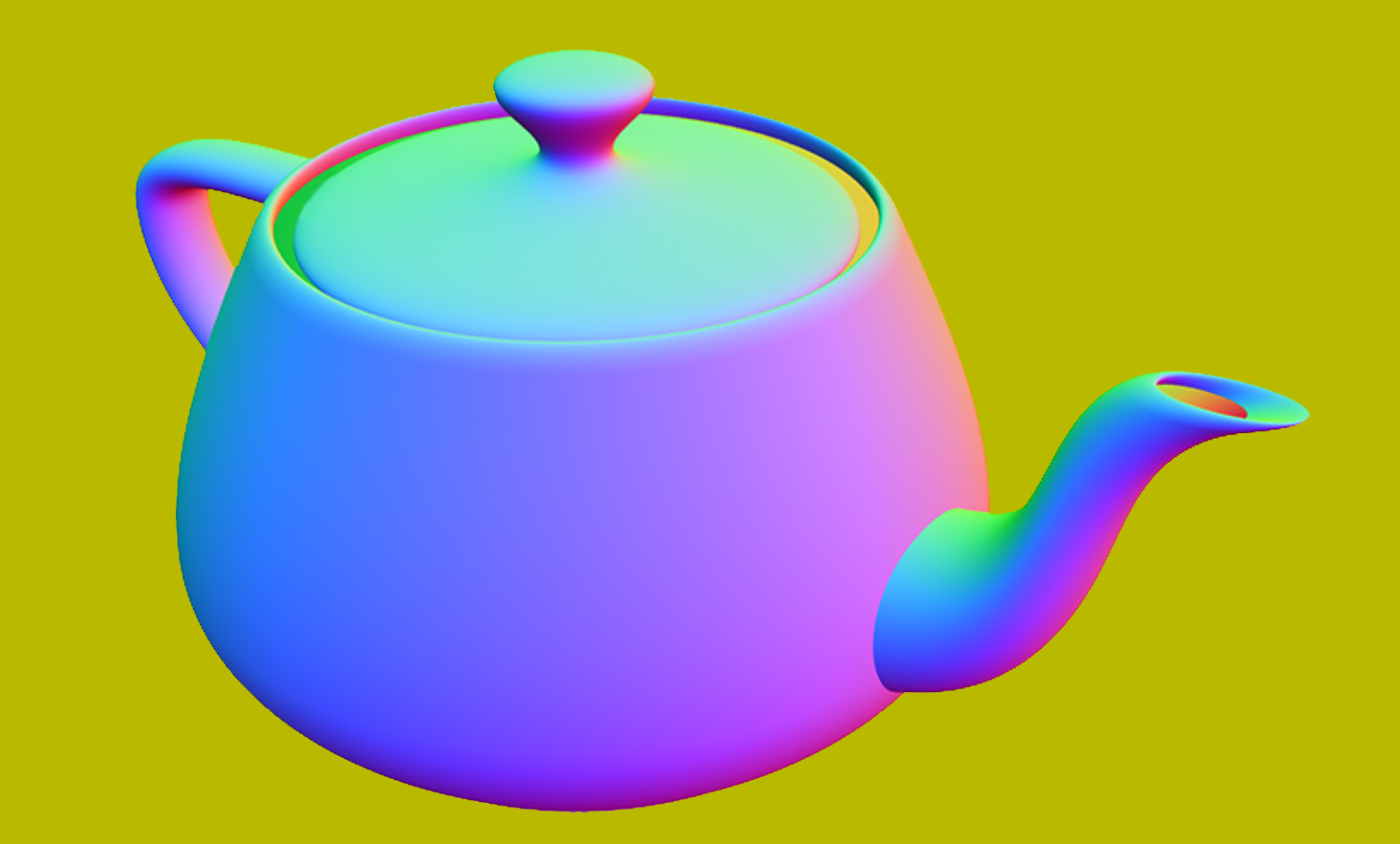}&
\includegraphics[width=0.45\textwidth]{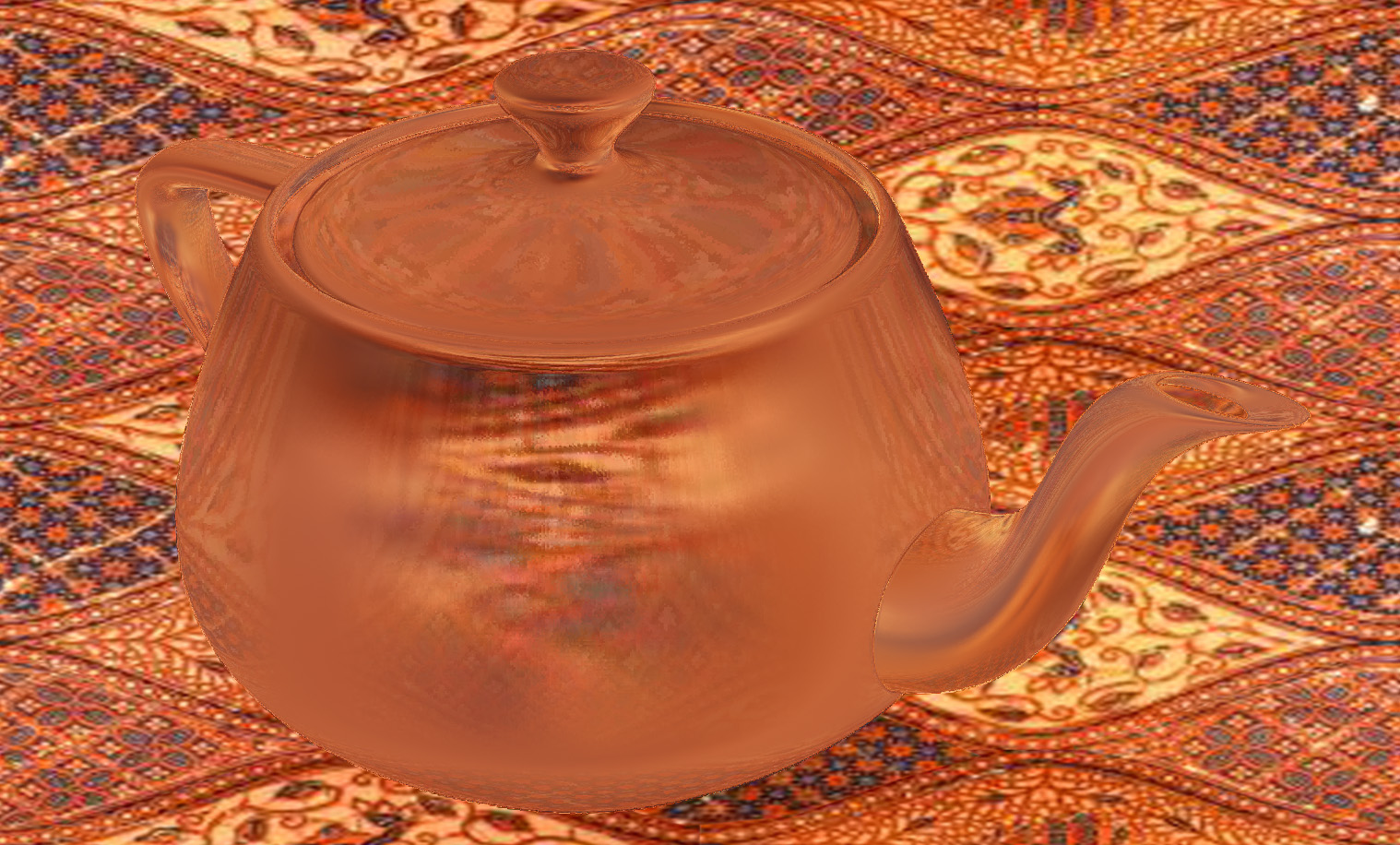}\\
Normal Maps &    Composite  \\
\end{tabular}
\end{center}
\caption{\it Reflection and refraction with Lumo and CrossShade models. In these cases, we do not have any foreground object, that is, $\alpha=0$. The composite images are simply the result of refraction and refraction of the same image that is used for both the environment and the background image. Lumo and CrossShade normal maps are used with permission.}
\label{fig_NormalMaps}
\end{figure}

\begin{figure*}[htb!]
\begin{center}
\begin{tabular}{cccccc}
\includegraphics[width=0.45\textwidth]{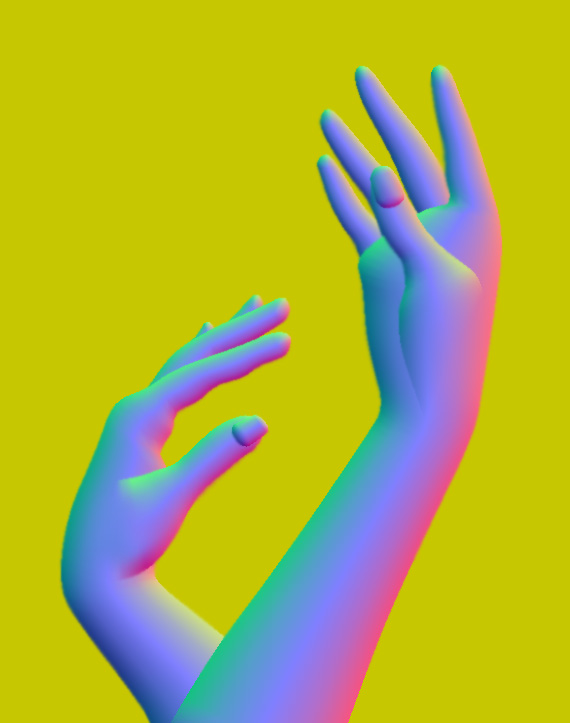}&
\includegraphics[width=0.45\textwidth]{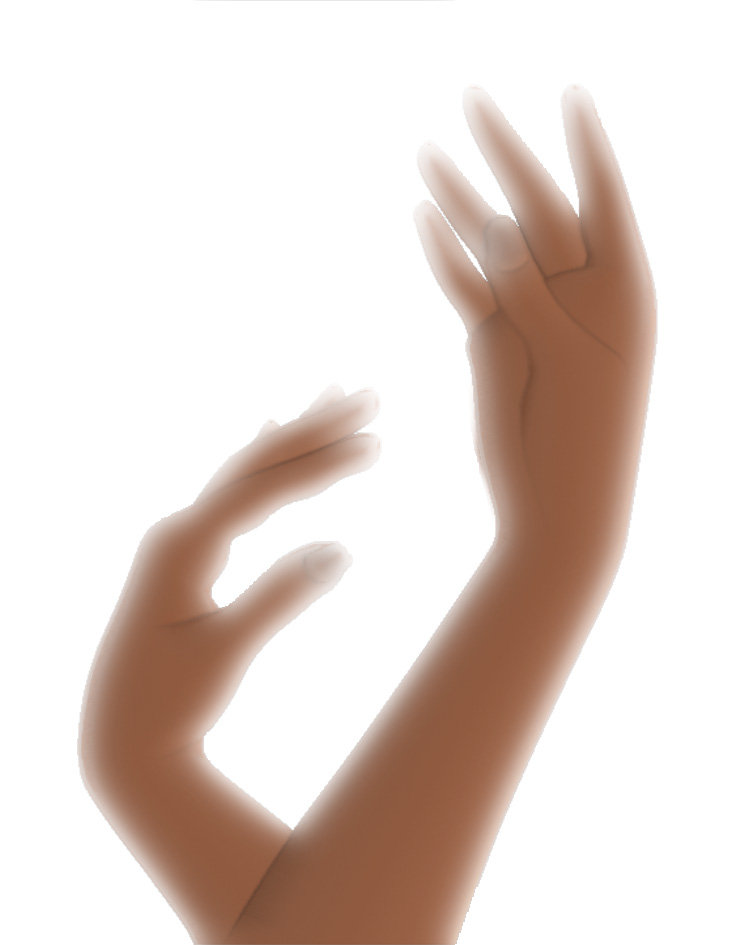}\\
(a) Shape Map & (b)  Foreground \\
\includegraphics[width=0.45\textwidth]{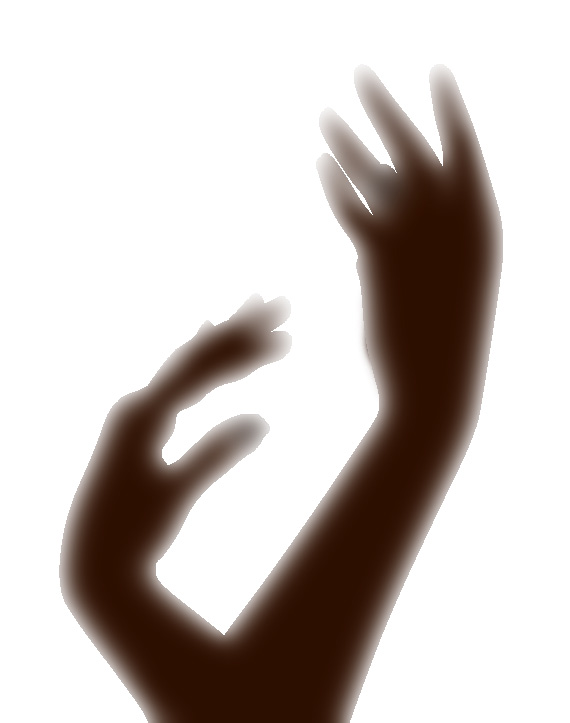}&
\includegraphics[width=0.45\textwidth]{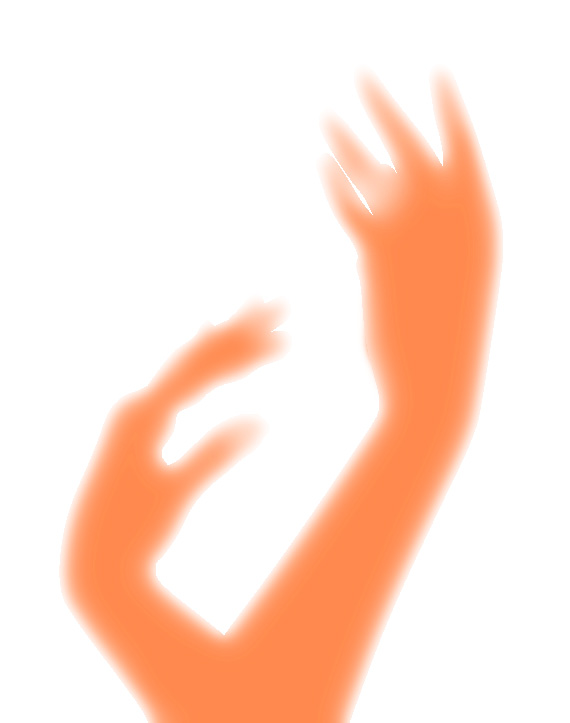}\\
(c) Dark Image & (d) Light Image \\
\end{tabular}
\end{center}
\caption{\it Subsurface scattering effect obtained by using $\alpha_FI$ term by using slight transparency around the silhouette regions of hands and arms. Note that the foreground image is slightly transparent in the thin regions. The foreground image is created using the method presented in \cite{akleman2024representing}. The shape map is also created with the sketch-based interface presented in \cite{akleman2024representing}. }
\label{fig_Hands}
\end{figure*}

Another method of obtaining shape maps is to photograph real objects using red and green lights, which can be a simple alternative to environmental matting \cite{Zongker1999}.  Figures~\ref{fig_photos0}, \ref{fig_photos1}, and~\ref{fig_photos2} show three such examples. In these examples, we have made only minimal changes in the original images: (1) we removed and replaced backgrounds with yellow color, and (2) we added a non-zero blue value for object regions. Note that the yellow background corresponds to zero thickness and therefore does not refract light. Although a constant thickness is not correct, the resulting refractions appear reasonably convincing. Artists can, of course, further manipulate these photographs to obtain desired effects.

\begin{figure*}[htb!]
\begin{center}
\begin{tabular}{cccccc}
\includegraphics[width=0.45\textwidth]{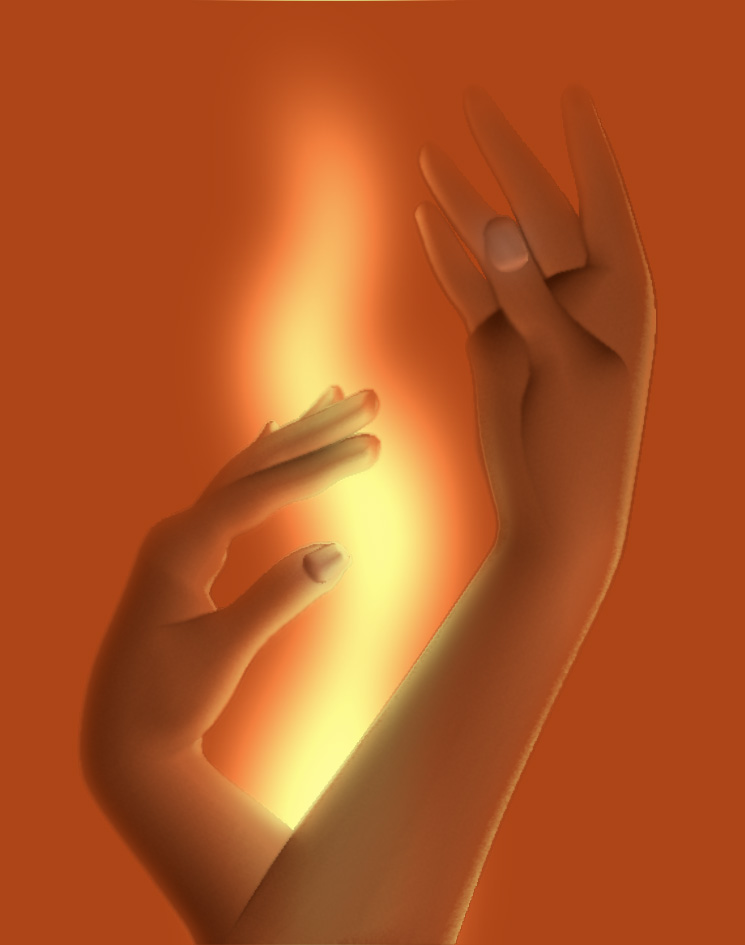}&
\includegraphics[width=0.45\textwidth]{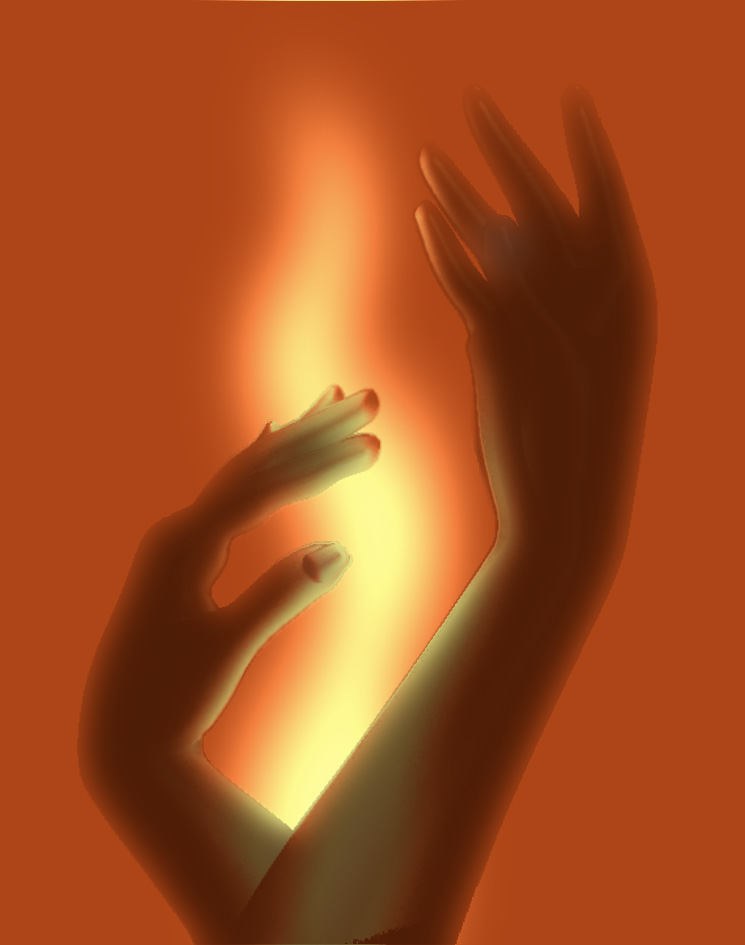}\\
Compositing 1 &  Compositing 2 \\
\end{tabular}
\includegraphics[width=0.45\textwidth]{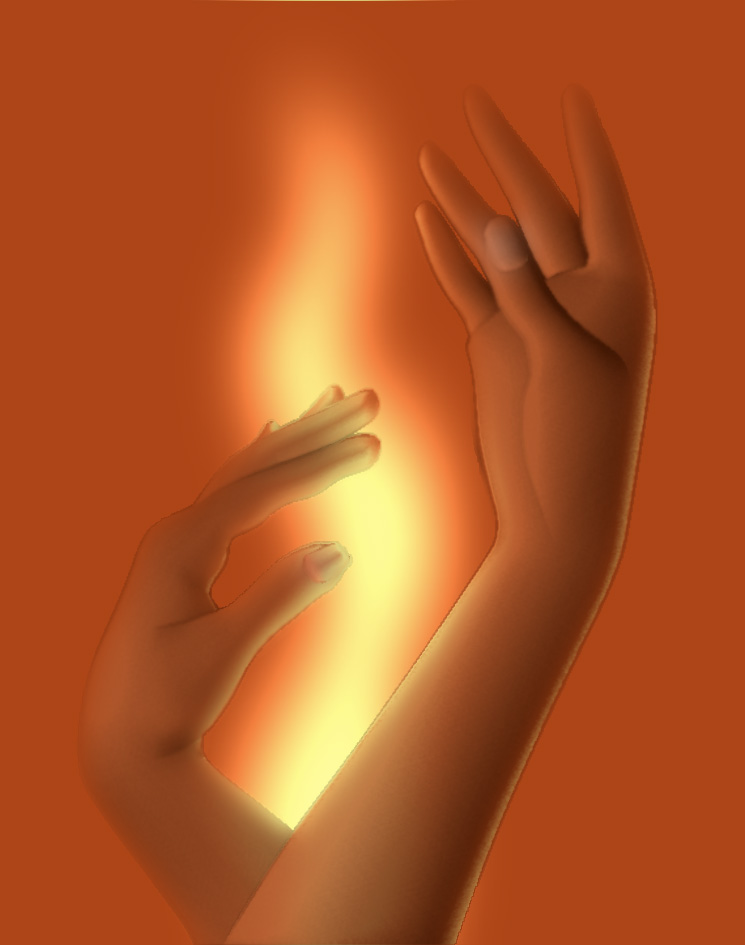}\\
Compositing 3
\end{center}
\caption{\it Subsurface scattering effect obtained by using $\alpha_FI$ term by using slight transparency around the silhouette regions of hands and arms. Note that the foreground image is slightly transparent in the thin regions. The shape map and the foreground image are created with our sketch-based interface. }
\label{fig_Hands2}
\end{figure*}

Another option, particularly for illustrators, is to model 2D vector fields directly with the sketch-based interface. To create 2D vector fields, we have implemented a simple modeling system based on boundary gradient interpolation, a concept suggested by Johnston \cite{Johnston2002}. For interpolation, we used a sketch-based modeling tool that uses subdivision \cite{akleman2024representing}. Since modeling is not the focus of this paper, we do not include details here (for details, see \cite{akleman2024representing}). Our rendering approach does not depend on our models. Normal maps created by Lumo \cite{Johnston2002} or CrossShade \cite{Shao2012} can be used as shape maps. Although neither of them provides separate depth information, as discussed earlier, the normal maps' blue channel can be used as depth if there is no other option. Figure~\ref{fig_NormalMaps} shows examples of shadow and reflection effects in the Lumo and CrossShade models.

\section{Compositing with Shape Maps \label{sec:compositing}}

Let $FI(u,v)$ be the image we want to composite. This particular image can refract and reflect images behind and in front of it. For simplicity, we assume that there is only one image behind and in front of the rendered shape-map image. Let $BI(u,v)$ and $EI(u,v)$ denote these images, called ``background image'' and ``environment image'', respectively, and let $\alpha_FI(u,v)$ denote the opacity of $FI$. This term is used to make only certain parts of the image transparent or reflective or to obtain the effect of subsurface scattering (see Figures~\ref{fig_Hands} and ~\ref{fig_Hands2}). We also have a user-defined global parameter, $\alpha_G$, such as in GIMP or Photoshop, that can make the entire layer more transparent. Then $\alpha_C(u,v) = \alpha_G \alpha_FI(u,v)$ denotes the combined transparency of the layer $FI(u,v)$. Using combined transparency, $\alpha$, we compute the composite image as follows: $$ CI (u, v) = \alpha FI + (1-\alpha) ( f EI(R) + (1-f) BI(T)).$$
where $R(x,y) = (u,v)$ represents reflection mapping; $T(x,y,\eta)=(u,v)$ represents refraction mapping; and the term $f(x,y,\eta)$ represents a Fresnel term. For perfect mirrors, we simply use a Fresnel term $f(x,y,\eta)=1$. For each of these functions, it is possible to use physically correct formulas, but those formulas do not provide the kind of artistic control that we want to provide. Therefore, we introduced linearized formulas that can provide artistically inspired versions of refraction, reflection, and Fresnel, which can be specifically used for compositing images.

\subsection{Reflection and Glossy Reflection \label{sec:reflection}}

For reflection mapping, our goal is to provide a simple and intuitive-to-use method for 2D artists. Our method is closely related to the mapping of spheres, which is one of the most widely used reflection methods \cite{Johnston2002}. In sphere mapping, a faraway spherical environment is stored as an image that depicts what a gazing ball (i.e. mirrored sphere) would reflect if it were placed into the environment, using an orthographic projection. Although spherical mapping is one of the simplest methods for obtaining reflections, it can still be complicated for some 2D artists who want to work only with rectangular images. Moreover, in our case, an additional problem comes from the fact that $x^2+y^2$ can be larger than $1$. Therefore, we need a similar method that can use rectangular environments. Fortunately, there exists a gazing ball shape that can reflect the whole environment into a square image. This particular gazing ball shape can be given by an implicit surface $max(x^2,y^2)+z^2=1$. We do not provide details here, but it can be shown that this shape, when placed in an environment, reflects the whole environment in a square using an orthographic projection along the $z$ direction. We can then create reflection simply by using the following operation:
\begin{equation} (u,v)=R(x,y)=(0.5x+0.5, 0.5y+0.5) \end{equation}
The only caveat in this approach is that every point on the boundary of a square image reflects the same point in the environment sphere. Therefore, every point on the boundary of the square image has to be the same. In practice, any seamlessly tileable wallpaper image can be used as an environment map. In our experience, any image works as an environment map, probably due to humans' high tolerance for discontinuities in mirror images.

For realistic reflections, we move the center of the environment image in tandem with the light position. This creates visually acceptable specular reflections. Therefore, we do not think that an additional specular highlight is necessary. Glossy reflection is simply obtained using smoothed versions of the environment maps provided by mipmap. Glossy reflections combined with other compositing operations such as multiplication can be used to obtain other effects such as environment illumination.

\subsection{Refraction and Translucency\label{sec:refraction}}

For refraction, we developed an artificial refraction mapping that provides simple and intuitive artistic control:
\begin{equation} (u,v)=T(x,y)= (u,v) - a \; d (x,y) \end{equation}
where $a \in [-1,1]$ is a user-defined global parameter that corresponds to the index of refraction as the value of $a$ is computed $log_2$ of $\eta = \eta_2/\eta_1$. Although shape maps are not supposed to have well-defined unit normal vectors, it is still possible to evaluate the physical sense of this equation qualitatively. For example, when $a=0$, there is no displacement of the background image, which is exactly what we expect when $\eta = 1$. In the regions where $d=0$ regardless of $a$, again there is no displacement, which is also what we expect to see in transparent regions that are extremely thin.

The values of $a$ that are higher than $0$ correspond to $\eta>1$, which is representative of the travel from air to water or glass. In this case, the incoming eye ray is $(0,0,-1)$ since we assume orthographic projection. The refracted ray will bend toward the vector $(-x,-y,-1)$. Therefore, we can assume that this vector can be given as a weighted average of the two vectors $(0,0,-1)$ and $(-x,-y,-1)$. Using $a$ as weight, we obtain a vector $(-ax,-ay,-1)$. If the object consists simply of two plates with thickness $d$, this vector will hit the back side of the object at $(u-d a x,v-d a y,-d)$. Once it hits the back side, we assume that it is refracted back to $(0,0,-1)$ and continues until it hits the background image. Therefore, we can simply approximate the refraction as a 2D displacement $(- d a x, -d a y )$.

The biggest advantage of this equation is that the refraction changes linearly with the global parameter $a$. This provides predictable control to artists. Following the same logic, the equation provides a vector between $(0,0,-1)$ and $(x,y,-1)$ for negative $a$ values. Although the length is not exactly correct, the direction of displacements is correct (this also provides predictable control). Translucency is obtained simply by using smoothed versions of the background images. Furthermore, we smoothed the background images according to the value of $|( d a x, d a y )|$ to improve visual quality and realism.

Gradient-domain refraction can also provide easy-to-understand and control deformations. It is also possible to obtain simpler operations such as translation, scaling, and rotation with refraction. For example, a 2D vector field $(x,y)=(v, -u)$ rotates a background image $45^{0}$ and uniformly scales the image by $\sqrt{2}$. From the same shape map, using the $a$ value, one can obtain a full range of scaled rotations. Shape maps that rotate are also good examples of impossible objects since they do not correspond to any 3D shape.

\subsection{Fresnel Effect \label{sec:FE}}

For artists, the Fresnel effect is important since it changes how reflection and refraction are composited on the basis of the incident angle and the index of refraction. From an art direction point of view, the number of parameters for defining Fresnel is undesirable. Therefore, the first issue is to reduce the number of parameters. Let $t=\theta(x,y)$ be a function that converts $(x,y)$ into a single variable between $[0,1]$ corresponding to the incident angle. On the basis of our earlier discussion, such a function can be obtained in several ways. Since there already exists a user-defined global parameter $a$ that provides the index of refraction, we can rewrite the variables of the Fresnel equation as $f(x,y,\eta) = f(t, a)$. By observing that the most important issue in Fresnel is to control the regions of strong reflection with the index of refraction, we further simplified the equation in the following form.
\begin{equation} f(t,a)  = (0.5 a - 0.5 )f(t,-1) + (0.5 a + 0.5 ) f(t,1) \end{equation}
Based on this equation, artists need only define two Fresnel curves that are given for $a=-1$ and $a=1$. We can then simply interpolate these curves using $a$. We also observed that the curves $f(t,-1)$ and $f(t,1)$ can simply be piecewise linear curves. Figure~\ref{fig_Bottle} shows a hand-drawn effect obtained with an exaggerated Fresnel function. In a dynamic version, the user can move the positions of the white regions with a slider that controls the value of $a$.


\section{Discussion, Conclusion and Future Work\label{sec:results}}  

Shape maps are particularly useful for 2D artists whose primary focus is painting and illustration. These artists have a good understanding of how reflection and refraction work, but may not want to follow the conventional 3D rendering process for obtaining 2D works. They also want to have creative control over the results. Shape maps allow them to create paintings that can be composited exactly as they want.  Our new compositing equation can be incorporated into existing 2D painting software without a major change. For simplicity of presentation, we introduced our compositing equation for only three layers; however, it is possible to extend it to many layers by introducing an eye layer that can divide layers into two categories front and back. The composited image is then obtained by refracting the layers in front of the eye and reflecting the layers behind the eye. As mentioned above, it can also be useful to extend Lumo \cite{Johnston2002} and CrossShade \cite{Shao2012} to provide control over the values of $d$.

\bibliographystyle{unsrtnat}
\bibliography{references}

\end{document}